\documentclass[9pt]{IEEEtran}
\usepackage{cite}
\usepackage{amsmath,amssymb,amsfonts}
\usepackage{algorithmic}
\usepackage{graphicx}
\usepackage{CJKutf8}
\usepackage[caption=false]{subfig}
\usepackage{easyReview}
\usepackage{hyperref}
\hypersetup{
    colorlinks=true,
    linkcolor=blue,
    citecolor=blue,
    urlcolor=blue
}
\begin{document}

%\title{{\fontsize{24}{26}\selectfont{Communication\rule{29.9pc}{0.5pt}}}\break\fontsize{16}{18}\selectfont Inverse Design Method Starting from Phaseless Radiation Pattern Based on Spectral Analysis} %\title{Inverse Design Method Starting from Phaseless Radiation Pattern Based on Spectral Analysis}
\title{\break\fontsize{16}{18}\selectfont Determining Aperture Field for Arbitrary Phaseless Far-Field Utilizing Inverse Design Method Based on Spectral Analysis}
\author
{
Chuan-Sheng Chen, Ren Wang, \IEEEmembership{Member, IEEE}, Jin-Pin Liu, and Bing-Zhong Wang, \IEEEmembership{Senior Member, IEEE}
\thanks
	{
	Manuscript received on June 14, 2023; revised Nov. 10, 2023. 
	This work was supported in part by the National Natural Science Foundation of China under Grants 62171081 and 61901086, the Natural Science Foundation of Sichuan Province under Grant 2022NSFSC0039. (\textit{Corresponding authors: Bing-Zhong Wang and Ren Wang.}) 
	}
\thanks
	{
	All four authors are with the Institute of Applied Physics, University of Electronic Science and Technology of China (UESTC), Chengdu, 611731, China,	(e-mail: cschen1997@gmail.com, rwang@uestc.edu.cn, jpinliu@gmail.com, and bzwang@uestc.edu.cn). 
	Ren Wang is also with the Yangtze Delta Region Institute (Huzhou), UESTC, Huzhou, China.
	}
}
\maketitle

\begin{abstract} 
In this communication, we propose an inverse design method based on spectral analysis  (IDMBSA) for achieving desired far-field radiation pattern through aperture field design. The aperture field obtained through IDMBSA can be utilized for array synthesis and sparse array design, and it can also serve as a design objective for existing aperture field implementation methods. 
IDMBSA combined with the coordinate transformation enables the fulfillment of design requirements for arbitrary phaseless radiation pattern and polarization needs. Non-linearity introduced by phaseless is addressed with a multi-objective optimization (MOO) algorithm. 
Compared to traditional array synthesis methods, IDMBSA significantly reduces the number of optimization variables by using modal expansion and further reduces the computational burden by utilizing analytical solutions. 
%Compared to existing methods, IDMBSA utilizes modal expansion to effectively reduce the number of optimization variables while ensuring the smoothness of the aperture field results, which is beneficial for physical implementation. By integrating the far-field asymptotic solution of the spectral domain method and employing analytic integration, IDMBSA successfully alleviates computational burden.
The inherent smoothness of the aperture field obtained through IDMBSA allows for direct application in existing aperture field implementation methods, facilitating the direct design of radiating devices with arbitrary far-field radiation pattern.
Simulations were conducted to explore IDMBSA's application in array synthesis, sparse arrays, and dual-polarization independent design, the results show the practicability of IDMBSA and its wide application prospect. 

\end{abstract}

\section{Introduction}\label{section1}
\IEEEPARstart{T}{he} radiation pattern is a crucial parameter in antenna systems and is a hot spot for scholars' research. Its requirements depend on the application scenarios and performance needs. 
%辐射方向图是电磁系统设计的关键参数，也是学者们的研究热点，辐射方向图的要求取决于具体的应用场景和性能需求。
Researchers have explored several approaches to shape the radiation pattern. 
One is to utilize array synthesis techniques to obtain the amplitude and phase information of each element's feeding signal, 
and then implement it through the array antenna\cite{bucci_1990, bucci_1994,bucci_1995,PSO_array_syh,bucci_GA,PSO848, ArraySynthesis_1_phaseonly,ArraySyhthesis_NumMore,IrregularArray}, typically with element spacing close to half a wavelength.
%主要是应用于间隔接近半波长的天线阵列. 
With the development of metamaterial,  array with sub-wavelength unit cell have gradually been widely used. 
%随着meta materials，metaantenna的迅速发展，使得亚波长单元阵列也逐渐得到了广泛的应用。
Researchers directly design antennas that meet specific requirements, such as transmit arrays, reflect arrays\cite{Transmit_Reflect_ArrayUsingSparse_Array_Method}, and metasurface antennas\cite{Metantennas, rew_2-4_add_TJCui}. 
Further, the device design methods which directly target the aperture field \cite{TensorMetasurfaceGetArbitraryAperture,CavityGetArbitraryAperture}, have gradually received attention from scholars. 
%更进一步的，直接以口径场为目标的器件设计\cite{TensorMetasurfaceGetArbitraryAperture} and \cite{CavityGetArbitraryAperture}，也逐渐收到了学者的关注。
%为了实现特定的辐射方向图，有两种常见的解决方案。一种是利用阵列综合技术获取各单元的馈电幅度和相位信息，然后通过相控阵天线实现。另一种是直接设计符合..特定要求的天线，如透射阵列、反射阵列天线、超表面天线等（引用）。

Among early efforts in array synthesis, several effective methodologies were introduced by \cite{bucci_1990, bucci_1994}, establishing the foundation for addressing array synthesis. Furthermore, Bucci and colleagues extended the work by using a generalized projection algorithm to reduce the possibility of the synthesis algorithm being trapped by spurious solutions in a subsequent publication \cite{bucci_1995}.  
%\cite{bucci_1990} presented an effective method for array pattern synthesis that allows a general synthesis problem to be dealt with.
%\cite{bucci_1995} utilize a generalized projection algorithm to overcome the possibility of the synthesis algorithm being trapped by spurious solutions while dealing with constraints on the excitation coefficients. 
With the increasing computational capabilities of computers, Multi-Objective Optimization (MOO) algorithms, such as Particle Swarm Optimization (PSO) \cite{MOO_PSO,MOO_PSO2} and Genetic Algorithms (GA)\cite{MOO_GA1} have found applications in array synthesis\cite{bucci_GA,PSO_array_syh,ArraySynthesis_1_phaseonly,PSO848}. 
%随着计算机计算能力的增加，以PSO、GA\cite{MOO_PSO,MOO_PSO2}为代表的多目标优化算法被应用与阵列综合中
Array synthesis techniques often require simultaneous optimization of both amplitudes and phases of elements, and sometimes only phase optimization is required \cite{ArraySynthesis_1_phaseonly}. 
This presents a significant challenge due to the notable increase in the dimensionality of the solution space as the number of elements grows \cite{ArraySyhthesis_NumMore,ArraySynthesis_1_phaseonly, PSO848}. 
%一个随之而来的挑战是随着单元数目的增加，解空间的维度也随之显著增加。
%一般的优化方法往往需要对单元的幅度和相位同时进行优化\cite{ArraySyhthesis_NumMore}，从而涉及到更多的待优化变量。
For instance, to optimize the feed phase for an array with the electrical size of $17\lambda$, \cite{PSO848} utilizes PSO to search solutions in the 848-dimensional solution hyperspace. 
%例如，为了电尺寸为$17\lambda$的阵列的馈电相位优化，文献\cite{PSO848}使用PSO方法在848维空间中寻找解，
\cite{IrregularArray} proposes a method to reduce the number of variables using irregular arrays, but the method is only applicable when maximizing directivity is desired. 
%文献\cite{IrregularArray}提出了一种利用非规则阵列减少待求解变量的方法，但该方法仅适用于追求最大方向性的情况。
Furthermore, due to the independence of optimization variables in array synthesis, there can be cases where adjacent elements have similar excitation amplitudes but opposite phases, which leads to increased discrepancies between simulation and measurement results. 
%此外，由于阵列综合中优化变量间是相互独立的，优化结果中会出现相邻单元激励相位突变的情况（幅度相近、相位相反），这会增加仿真与实测结果之间的差异（尤其是对于单元间耦合较大的阵列）。

The popularization of arrays featuring sub-wavelength unit cells, such as metaantennas, has drawn the focus of scholars toward novel methodologies.
%随着array with sub-wavelength unit cell的推广，一些新的方法被学者所注意。应用
%In the process of directly designing specific antennas, it is common to handle each case individually\cite{MethodForSpecificAntenna_1,MethodForSpecificAntenna_3}. 
%\remove{According to the authors’ knowledge, there is currently a lack of general design methods.}
%在直接设计特定天线的过程中，通常是根据具体情况采取特定的方法，据作者了解，目前缺乏通用的设计方法。（引用） 
\cite{convex_SDR} introduces a procedure for solving non-convex array synthesis problems, which is based on the semidefinite relaxation (SDR) technique. 
\cite{rew_2-4_add_TJCui} addresses the smooth-phase constraint to reduce the influence of the aperiodic arrangement of elements, and utilizes the alternating direction method of multipliers (ADMM) to achieve the synthesis of the Huygens metasurface. 
%为了 reduce the influence of the aperiodic arrangement of meta-atoms.在\cite{rew_2-4_add_TJCui} 考虑 addresses the smooth-phase constraint 的同时 借助了alternating direction method of multipliers (ADMM)实现了对 Huygens metasurface 的synthesis and Verification.

With the rise of metamaterial and metaantennas, scholars have shown interest in integrating array synthesis techniques with metasurface design to achieve a one-stop design. For example, in \cite{review2-4-1}, designers solve for the surface parameters based on their application-specific far-field criteria. 
%事实上随着超表面天线研究的兴起，学者们对将阵列综合技术与超表面设计结合，实现一站式设计产生兴趣,例如在\cite{review2-4-1}中designers solve for the surface parameters based on their application-specific far-field criteria. 
Furthermore, \cite{TensorMetasurfaceGetArbitraryAperture} focuses on the design of tensorial modulated metasurfaces able to implement arbitrary aperture field distributions. \cite{CavityGetArbitraryAperture} presents a design procedure for cavity-excited omega-bianisotropic metasurfaces generating desirable field distribution on their aperture. 
%值得注意的是，在文献\cite{TensorMetasurfaceGetArbitraryAperture}提出了一种利用张量超表面实现任意口径场的方法，其中特别指出了平滑的阻抗变换可以有效避免不必要的反射，从而提高设计精度。 

\cite{TensorMetasurfaceGetArbitraryAperture} and \cite{CavityGetArbitraryAperture} demonstrate the feasibility of achieving arbitrary aperture field and emphasize the significance of continuous aperture fields. 
%\cite{TensorMetasurfaceGetArbitraryAperture}和\cite{CavityGetArbitraryAperture} 的两篇文章阐述了实现任意口径场分布的可行性，并强调了连续的口径场的重要性。然而，它们并没有深入研究如何选择或设计特定的口径场分布。没有讨论如何按需设计口径场“they do not delve into the investigation of selecting or designing specific aperture field distributions.
%如果将电磁场的应用粗略的划分为远场和近场应用两类，那么由远场、近场获得口径场的方法是不可或缺的。
The methods to obtain aperture fields from the far-field and near-field are indispensable. However, \cite{TensorMetasurfaceGetArbitraryAperture} and \cite{CavityGetArbitraryAperture} do not discuss how to design the aperture field on demand. The authors of \cite{TensorMetasurfaceGetArbitraryAperture} put forward a method to derive the aperture fields generating the required near-field radiation in \cite{TensorMeta_NearFieldGetAmpture}. Similarly, we presented a method for converting port field distribution into a device's internal field distribution using time-reversal techniques in \cite{self_TRconversion}. 
%在\cite{TensorMeta_NearFieldGetAmpture}中，作者介绍了一种通过从近场分布反向传播获得口径场的方法。同样地，在\cite{self_TRconversion}中，我提出了一种利用时间反演技术从近场获得口径场的方法。
% add

This communication proposes an inverse design method based on spectral analysis (IDMBSA)
to determine aperture field from arbitrary phaseless far-field radiation pattern (phaseless but with polarization information). 
%IDMBSA is frequency-independent and allows for arbitrary dimensions of the desired aperture field, demonstrating its excellent versatility. 
IDMBSA has broad prospects for application, as it can be utilized to provide on-demand aperture field distribution for physical implementation methods like \cite{TensorMetasurfaceGetArbitraryAperture} and \cite{CavityGetArbitraryAperture}. It naturally adapts to the polarization constraint and also can be used to overcome the drawback of excessive optimization variables in existing array synthesis methods such as \cite{ArraySynthesis_1_phaseonly,PSO848,ArraySyhthesis_NumMore}. 
IDMBSA is also applicable to sparse array design and can independently design the radiation patterns for different polarizations. 
%与传统的功率综合方法相比，IDMBSA又内禀的可以引入远场的极化信息

%为克服上述现有的阵列综合中方法优化变量过多的弊端，为[4]和[5] 这类口径场物理实现方法提供按需设计的口径场.本文提出了一种通用的由远场功率方向图（考虑极化）得到口径场分布的逆设计方法(IDMBSA)，方法与频率无关，待求口径场的尺寸也可以任意给定，具有良好的通用性。 IDMBSA相比与已有的基于EFIE的逆设计方法\cite{inverse_MP_ontheuse}，通过结合谱域法的远场渐近特性来降低计算成本，并通过与坐标变换相结合来自然地适应不同的极化约束。利用模式分解将口径场展开为模式系数的线性组合，通过将模式系数作为待求解变量，在大幅度减少解空间的维度、降低问题复杂性的同时，也保证了场的平滑性。确定的模式项使利用解析积分降低计算负担成为可能。最终借助多目标优化算法完成对模式系数的求解，也即完成了对口径场的求解。
Specifically, by integrating the far-field asymptotic solution of the spectral domain method and employing analytic integration, IDMBSA successfully alleviates computational burden compared to \cite{inverse_MP_ontheuse}. IDMBSA utilizes modal expansion \cite{collin1990field} to effectively reduce the number of optimization variables while ensuring the smoothness of the aperture field results, which is beneficial for physical implementation. Lastly, PSO \cite{MOO_PSO,MOO_PSO2} is employed to determine the modal coefficients and obtain the aperture field. Additionally, IDMBSA combined with the \textit{Ludwig3} coordinate transformation \cite{ludwig3} enables the fulfillment of design requirements for arbitrary radiation pattern and polarization needs. 

%In summary, our proposed IDMBSA demonstrates efficiency, versatility, and accuracy in solving the inverse design problem of obtaining aperture field distribution from phaseless radiation pattern. It overcomes the limitations of traditional array synthesis methods, facilitates the design of sparse arrays, and enables more continuous target aperture fields for physical implementation methods. Thus, IDMBSA holds significant potential for various applications.
%我们的IDMBSA方法在从无相辐射图中获取口径场分布的反设计问题中表现出高效率、通用性和准确性。它不仅克服了传统阵列综合方法的限制，还为稀疏阵列设计提供了便利，并使超表面天线的目标孔径场更加连续。因此，IDMBSA具有广泛的应用潜力。

The rest of this communication is presented as follows. Section II provides the background of the problem. Section III presents the inverse design method, including the specific derivation and solution process. 
In section IV, numerical simulations of aperture field obtained using IDMBSA are given, and verified by full-wave simulation experiments. In Section V, we present multiple cases to demonstrate the practicality of IDMBSA in array synthesis, sparse arrays, independent design for two polarizations, and so on.

\section{Background}\label{section2}
\begin{figure}[ht]
\vspace{-0.8em}
	\centerline{\includegraphics[width=0.5\columnwidth]{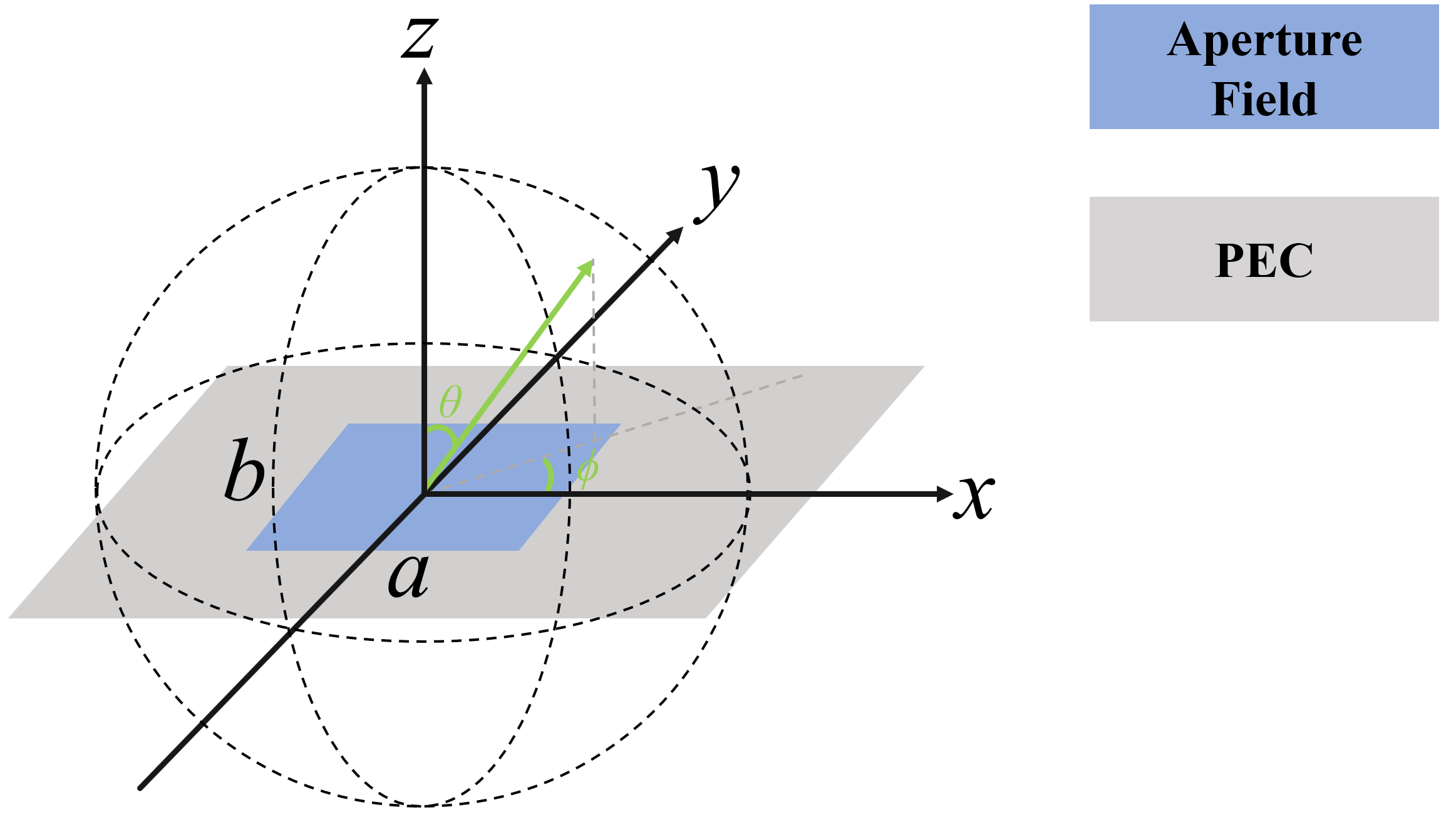}}
	\caption{Schematic diagram of the general forward aperture problem }
	\label{fig1}
\end{figure}

The general aperture radiation problem is calculating the radiation field from the known aperture field $ \boldsymbol{E}_{\mathrm{ape}}(x,y)|_{z=0} $. 
Its schematic diagram is shown in Fig. \ref{fig1}. 
The analysis below consider only the upper half-space ($z>0$), that is $\theta \in [0,{\pi}/{2}]$. 
The aperture field $\boldsymbol{E}_{\mathrm{ape}}$, defined over the domain $\left\lbrace (x, y) \,|\, x \in [-\frac{a}{2}, \frac{a}{2}], y \in [-\frac{b}{2}, \frac{b}{2}] \right\rbrace$, is mounted on an infinite ground plane (PEC) located at $z = 0$.
According to the uniqueness theorem, as long as the tangential components of $ \boldsymbol{E}_{\mathrm{ape}}(x,y) $ ($ E_{\mathrm{ape},x} $ and $ E_{\mathrm{ape},y} $) are known, $\boldsymbol{E}(x, y, z )|_{z >0}$  is uniquely determined. 
This forward problem can be analyzed in spatial or spectral domains. 

%In the spatial domain analysis, it can be transformed into a problem of magnetic current $(\boldsymbol{M}_s = -2 \boldsymbol{n}\times \boldsymbol{E}_a) $ radiation in free space\cite{536088}. $\boldsymbol{E}(x, y, z )|_{z >0}$ can be calculated by using the Green's function method shown in \cite{inverse_MP_ontheuse} or by introducing the auxiliary potential functions $\boldsymbol{A}$ and $\boldsymbol{F}$. 

In the spectral domain, as written in \eqref{eq2}, the monochromatic wave field $\boldsymbol{E}(x, y, z )$ radiated by the aperture field can be regarded as a superposition of plane waves of the form $\boldsymbol{f}\left( \theta, \phi \right) \exp({-j\boldsymbol{k}\cdot\boldsymbol{r}})$\cite{8466116}. 
\begin{equation}\label{eq2}
\begin{aligned}
	& \boldsymbol{E}(x,y,z)=\frac{1}{4\pi ^2}\int_{-\infty}^{\infty}{\int_{-\infty}^{\infty}{\boldsymbol{f}\left( \theta, \phi \right) e^{-j\boldsymbol{k}\cdot \boldsymbol{r}}}}\mathrm{d}k_x\mathrm{d}k_y 
\end{aligned}
\end{equation}
where
$ k_x = k \sin \theta \cos \phi $, $ k_y = k \sin \theta \sin \phi$, and $k_z^2 = k^2 - k_x^2 -k_y^2 $. 
 $ \boldsymbol{f} $ consists of $ f_x $ and $ f_y $ given by \eqref{eq-core}. 
%\begin{small}
\begin{subequations}\label{eq-core}
\small
\begin{align}
	f_x(\theta,\phi)&=\int_{-\frac{b}{2}}^{\frac{b}{2}}\int_{-\frac{a}{2}}^{\frac{a}{2}} E_{\mathrm{ape},x}(x', y') \cdot e^{j(k_x x' + k_y y')}\mathrm{d}x'\mathrm{d}y'	\label{eq-core-a}
	\\
	f_y(\theta,\phi)&=\int_{-\frac{b}{2}}^{\frac{b}{2}}\int_{-\frac{a}{2}}^{\frac{a}{2}} E_{\mathrm{ape},y}(x', y') \cdot e^{j(k_x x' + k_y y')}\mathrm{d}x'\mathrm{d}y'	\label{eq-core-b}
\end{align}\normalsize
\end{subequations}
%\end{small}
The far-field asymptotic evaluation of \eqref{eq2} is given by \eqref{eq_asym}, by using the \textit{Stationary Phase} method\cite{balanis2016antenna}. 
\begin{subequations}\label{eq_asym}
\begin{equation}\label{eq_asym_a}
	\boldsymbol{E}(r,\theta,\phi)  \approx\   j\frac{ke^{-jkr}}{2\pi r} \boldsymbol{e}_{\mathrm{far}}(\theta, \phi)
\end{equation}
\begin{equation}\label{eq_asym_b}
\begin{aligned}
	\boldsymbol{e}_{\mathrm{far}}(\theta, \phi) & = \boldsymbol {\hat{a}}_{\theta} e_{\mathrm{far},\theta}(\theta, \phi) + \boldsymbol{\hat{a}}_{\phi}e_{\mathrm{far},\phi}(\theta, \phi)
	\\
		& = \boldsymbol {\hat{a}}_{\theta} \left(f_x\cos\phi+f_y\sin\phi\right) 
	\\
 		& \quad + \boldsymbol{\hat{a}}_{\phi}[\cos\theta\left(-f_x\sin\phi+f_y\cos\phi\right) ]
\end{aligned}
\end{equation}
\end{subequations}

In most cases, we pay more attention to the pattern $ \boldsymbol{e}_{\mathrm{far}}(\theta, \phi) $ than the coefficient on distance $  j\frac{ke^{-jkr}}{2\pi r} $. Obviously, $ \boldsymbol{e}_\mathrm{far} $ here includes both polarization and phase information. 

\section{Inverse Design Method}
This section introduces the formulation of the inverse design method. Specifically, IDMBSA can be summarized into two steps. 
The first step is getting $ |f_x| $ and $ |f_y| $ from the desired $|e_{\mathrm{far},\theta}|$ and $|e_{\mathrm{far},\phi}|$.
Additionally, for practical engineering applications, we also provide formulas to extract $ |f_x| $ and $ |f_y| $ from other  polarization decompositions, such as horizontal (H) $ |e_{\mathrm{far},h}| $ and vertical (V) polarization $ |e_{\mathrm{far},v}| $. 
The second step is finding a suitable set of solutions ($ E_{\mathrm{ape},x} $ and $ E_{\mathrm{ape},y} $)  from $ |f_x| $ and $ |f_y| $.

%第一小节主要讲述无相位方向图的引入已有推导，第二小节主要讲述具体地求解方法。
%\subsection{Dealing with Phaseless Radiation Pattern}
\subsection{Dealing with Phaseless Radiation Pattern}
This subsection focuses on how to introduce phaseless patterns into the existing formulations, so as to convert phaseless pattern targets into constraints on aperture field.  % 本章主要讲述如何将无相位引入到现有的公式体系中，从而将远场目标转换为对口经场的约束。 
% as we said in Section \ref{section1}, exploring the \textit{cause} based on the \textit{effect}. 
%Using \eqref{eq_asym_b} can easily get $ \boldsymbol{e}_\mathrm{far} $ from $ \boldsymbol{f} $.
%After simple transformations, $ \boldsymbol{f} $ can also be obtained from $ \boldsymbol{e}_\mathrm{far} $ as shown in \eqref{eq6}. 
From \eqref{eq_asym_b} we can derive \eqref{eq6}, through algebraic transformations. 
\begin{subequations}\label{eq6}
\begin{align}
	f_x(\theta, \phi) &= e_{\mathrm{far},\theta}(\theta, \phi)\cos \theta -e_{\mathrm{far},\phi}(\theta, \phi) {\sin\phi}/{\cos\theta}
	\\
	f_y(\theta, \phi) &= e_{\mathrm{far},\theta}(\theta, \phi) \sin\phi + e_{\mathrm{far},\phi}(\theta, \phi) {\cos\phi}/{\cos\theta} 
\end{align}
\end{subequations}
%This is because one is more interested in sufficiently intense/low fields in the different directions. 
%但是，我们可已通过取模值的操作，将\eqref{eq6}变换为可以带入phaseless的形式，如公式7所示
During the inverse design process, what usually given is $ \left|e_{\mathrm{far},\theta}\right| $ and $ \left|e_{\mathrm{far},\phi}\right| $ (the absolute value symbol here represents phaseless).
Since $ \left|e_{\mathrm{far},\theta}\right| $ and $ \left|e_{\mathrm{far},\phi}\right| $  cannot be directly substituted into \eqref{eq6}, we take the modulus of both sides of \eqref{eq6} and obtain \eqref{eq_module}.
%The trigonometric function $ \cos$ and $ \sin $ in \eqref{eq_module} are known real number. 
%and its sign symbol can be determined after moving outside the modulus operator.
$ \gamma $ in \eqref{eq_module} is a function of $ \theta $ and $ \phi $, 
which represents the phase difference of two polarization components at a certain direction $ (\theta,\phi) $. 
%公式6中的三角函数是关于theta和phi的系数，其提出取模操作后的符号可以被确定；
% 公式6中的gamma也是关于a和b的函数，表示的是e的夹角
\begin{subequations}\label{eq_module}
\begin{align}
\begin{split}
	\left|f_x(\theta ,\phi )\right|=\bigg(&\left| e_{\mathrm{far},\theta} \right|^2\left| \cos \theta \right|^2+\left| e_{\mathrm{far},\phi}\frac{\sin \phi}{\cos \theta} \right|^2
	\\
	&\qquad -2\left| e_{\mathrm{far},\theta} \right|\left| e_{\mathrm{far},\phi} \right|\left| \sin \phi \right|\cos \gamma \bigg)^{\frac{1}{2}}
\end{split}
\\
\begin{split}
	\left| f_y(\theta ,\phi ) \right|=\bigg(&\left| e_{\mathrm{far},\theta} \right|^2\left| \sin \phi \right|^2+\left| e_{\mathrm{far},\phi} \right|^2\left| \frac{\cos\phi}{\cos \theta} \right|^2
	\\
	&\qquad +2\left| e_{\mathrm{far},\theta} \right|\left| e_{\mathrm{far},\phi} \right|\left| \frac{\sin \phi \cos \phi}{\cos \theta} \right|\cos \gamma \bigg) ^{\frac{1}{2}}
\end{split}
\end{align}
\end{subequations}
Combined with engineering applications, 
%Consider the given in engineering applications are ueually horizontal and vertical components, 
we also give another expression of \eqref{eq_module} for horizontal (H) and vertical (V) polarization ($ |e_{\mathrm{far},h}| $ and $ |e_{\mathrm{far},v}| $), by using the \textit{Ludwig3} coordinate system, as shown in  \eqref{eq_module_hv}. 
\begin{subequations}\label{eq_module_hv}
\begin{align}
&|f_x(\theta ,\phi )| = \sqrt{|l_1|^2 + |l_2|^2 + 2|l_1||l_2|\cos\gamma}
\\
&|f_y(\theta ,\phi )| = \sqrt{|l_3|^2 + |l_4|^2 + 2|l_3||l_4|\cos\gamma}
\end{align}
\end{subequations}
where 
\begin{equation*}
\begin{aligned}
l_1&=e_{\mathrm{far},h}\left( \cos ^2\phi +\sin ^2\phi /\cos \theta \right) ;\\
l_2&=e_{\mathrm{far},v}\sin \phi \cos \phi \left( 1-1/\cos \theta \right) ;\\
l_3&=e_{\mathrm{far},h}\sin \phi \cos \phi \left( 1-1/\cos \theta \right) ;\\
l_4&=e_{\mathrm{far},v}\left( \sin ^2\phi +\cos ^2\phi /\cos \theta \right) ;
\end{aligned}
\end{equation*}
%\end{footnotesize}
Several common cases regarding \eqref{eq_module_hv} are discussed next. 
%The subsequent discussions will also focus on \eqref{eq_module_hv}, we discuss several practical cases. 
\begin{itemize}
\item Case I: Linear Polarization

When only a single polarization is needed, another orthogonal polarization can be set to 0. 
For example, when only the V polarization $ e_{\mathrm{far},v} $ is considered, \eqref{eq_module_hv} degenerates to \eqref{eq_linear}.
\end{itemize}
% 例如仅考虑ev分量，则公式7退化为xxx
\begin{equation}\label{eq_linear}
|f_x(\theta ,\phi )| = |l_2|;\ |f_y(\theta ,\phi )| = |l_4|
\end{equation}
\begin{itemize}
\item Case II: Circular Polarization

Circular polarization requires that the modulus values of the two polarization components are equal, and the phase difference is $90^ \circ$. That is $ |e_{\mathrm{far},h}| = |e_{\mathrm{far},v}|$, and $ \gamma = 90^\circ $ in any direction $ (\theta,\phi ) $. Then, \eqref{eq_module_hv} degenerates to \eqref{eq_circular}. 
%The setting for $ \gamma $ is perhaps too strict, but still suitable as a target. 
\end{itemize}
\begin{equation}\label{eq_circular}
|f_x(\theta ,\phi )| = \sqrt{|l_1|^2 + |l_2|^2 };\ |f_y(\theta ,\phi )| = \sqrt{|l_3|^2 + |l_4|^2 }
\end{equation}
\begin{itemize}
\item Case III: Elliptical Polarization

Once $ \gamma $ is determined, \eqref{eq_module_hv} itself corresponds to elliptical polarization. 
\end{itemize}
\noindent The $|f_x|$ and $|f_y|$ in \eqref{eq_module} and \eqref{eq_module_hv} act as bridges to deliver the phaseless pattern constraints to the aperture field constraints. 

Obviously, it is easy to get $ |f_x| $ and $ |f_y| $ from the desired $|e_{\mathrm{far},\theta}|$ and $|e_{\mathrm{far},\phi}|$ or $|e_{\mathrm{far},h}|$ and $|e_{\mathrm{far},v}|$ by using \eqref{eq_module} or \eqref{eq_module_hv}. 

\subsection{Solving Integral Equation for Aperture Field}
Next, what we need is to find a suitable set of solutions ($ E_{\mathrm{ape},x} $ and $ E_{\mathrm{ape},y} $)  from $ |f_x| $ and $ |f_y| $  according to \eqref{eq_core_abs}.
\begin{subequations}\label{eq_core_abs}
\footnotesize
\begin{align}
	|f_x(\theta,\phi)|&=\left|\int_{-\frac{b}{2}}^{\frac{b}{2}}\int_{-\frac{a}{2}}^{\frac{a}{2}} E_{\mathrm{ape},x}(x', y') \cdot e^{j(k_x x' + k_y y')}\mathrm{d}x'\mathrm{d}y'\right|	
	\\
	|f_y(\theta,\phi)|&=\left|\int_{-\frac{b}{2}}^{\frac{b}{2}}\int_{-\frac{a}{2}}^{\frac{a}{2}} E_{\mathrm{ape},y}(x', y') \cdot e^{j(k_x x' + k_y y')}\mathrm{d}x'\mathrm{d}y'\right|	
\end{align}
\normalsize
\end{subequations}

During this step, the inverse Fourier transform cannot be applied in \eqref{eq_core_abs}, because the addition of the absolute value symbol on the right-hand side eliminates all phase information.
The root cause is the non-linearity introduced by the phaseless. (In mathematics, taking the modulus value on both sides of \eqref{eq-core} is \eqref{eq_core_abs}). 
Therefore, we can only treat \eqref{eq_core_abs} as the first kind of two-dimensional Fredholm integral equations after the modulo operation, which is difficult to obtain analytical solutions\cite{FredholmOverview}.
Here, we will employ modal expansion and MOO to approximately solve \eqref{eq_core_abs}.
%这里，我们将借助模式展开和多目标优化算法完成对\eqref{eq_core_abs}的近似求解。
\subsubsection{Modal Expansion on Aperture Field}\ 

First, expand the unknown integrand $ \boldsymbol{E}_\mathrm{ape} $  using the modal expansion method. The physical constraints can be cleverly introduced into the integral equation with the help of boundary conditions. As can be seen from Fig. \ref{fig1}, $ \boldsymbol{E}_{\mathrm{ape}} $ satisfies \eqref{eq_boundary}. 
The modal expansion of $ \boldsymbol{E}_{\mathrm{ape}} $ is expressed by the terms in the braces of \eqref{eq10}, and it follows easily that $ \boldsymbol{E}_\mathrm{ape} $ can be approximated by a finite number of unknown modal coefficients and known modal terms. 
\begin{equation}\label{eq_boundary}
\left\{
\begin{array}{l}
	E_{\mathrm{ape},x}|_{y=-\frac{b}{2},\frac{b}{2}}=0\\
	E_{\mathrm{ape},y}|_{x=-\frac{a}{2},\frac{a}{2}}=0\\
\end{array}
\right. 
\end{equation}

\begin{figure}[htb]
	\centering
	\includegraphics[width=0.8\columnwidth]{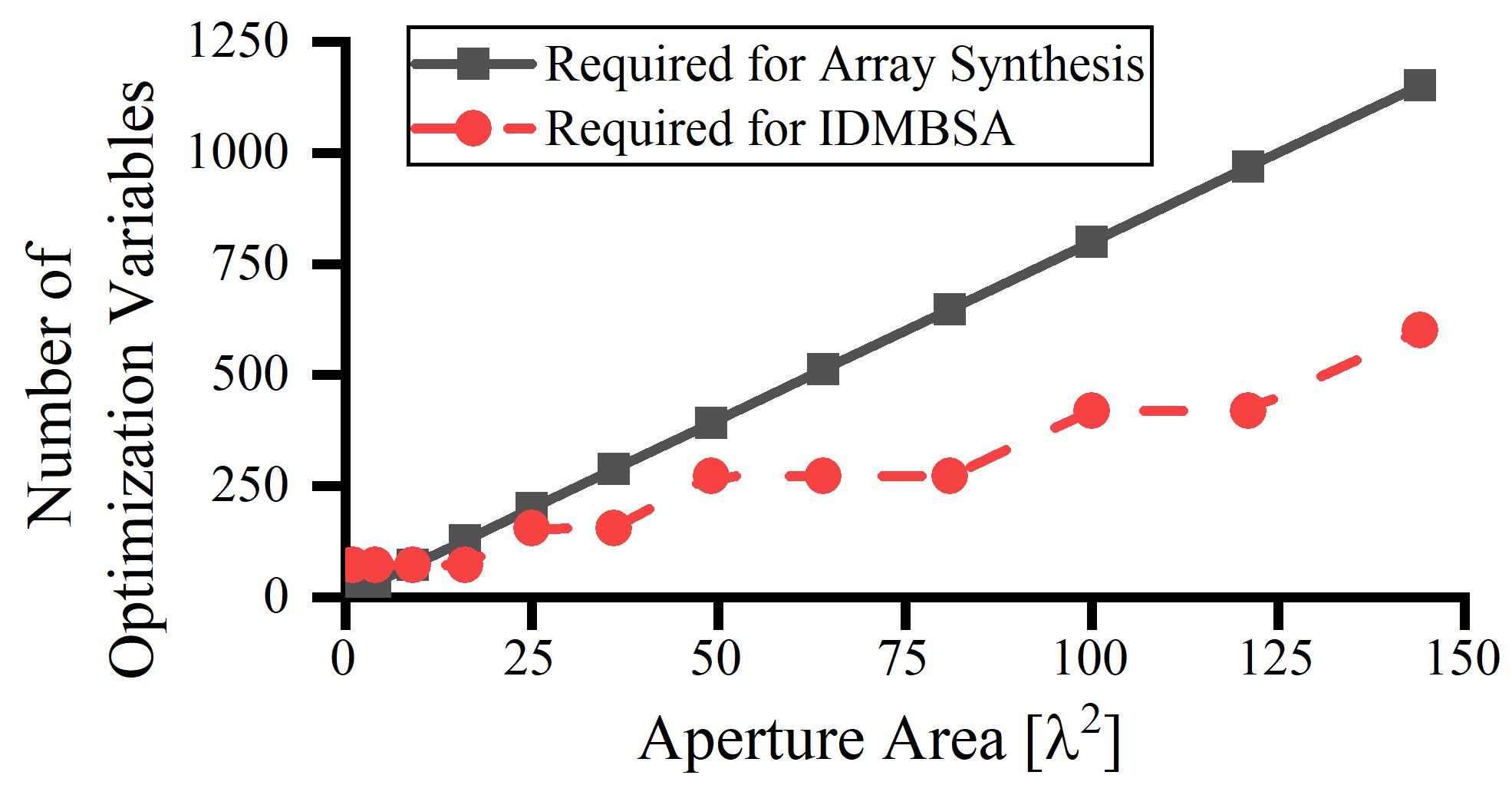}
	\caption{Variation of the number of optimization variables with aperture area}
	\label{fig_OptVarNum}
\end{figure}

Equation \eqref{eq10} converts the problem of solving $ \boldsymbol{E}_{\mathrm{ape}}(x,y) $ to solving the unknown complex modal coefficients  $\alpha_{mn}^{i}$ and $\beta_{mn}^{i}$, $(i=x,y)$. 
The selection of the basis function is not unique. The trigonometric functions were chosen because they are the harmonic functions of the waveguide harmonic equation, if we regard the modal shown in Fig. \ref{fig1} as a waveguide connected to the infinite PEC plane. 
%In addition, this selection is naturally consistent with the boundary conditions （在未来加快求解的过程中可能是十分有益的，不提及pinn）, which would be useful if this method could be migrated to PINN in the future}
The analytical integration of \eqref{eq10} is provided in Appendix \ref{sec:appd-a}. 
Thanks to the availability of an analytical expression, \eqref{eq10} can be integrated analytically instead of numerically, resulting in a significant improvement in computational efficiency.
The number of optimization variables in general array synthesis methods is proportional to the aperture area. %  is twice the number of ports and
For example, for an array with area {$L ^2 $} [$\lambda^2$], the number of ports is about $(L/0.5)^2 = 4L^2 $, if the spacing of elements is $0.5 [\lambda]$. The variable to be optimized is about $8L^2 $ (considering amplitude and phase). 
% 例如对于 面积为S($\lambda$^2)的阵列，若阵元间距为0.5lambda，其端口数量约为4L^2，对于幅、相优化，其待优化变量约为8L^2;而在IDMBSA中，只于模式的数目有关，
Whereas, in IDMBSA, the number of optimization variables is associated with the number of modes and increases slowly with the expansion of the aperture area. The determination of the number of modes in IDMBSA is elaborated in Section \ref{sec:sec4B}.
A rough comparison graph is summarized in Fig. \ref{fig_OptVarNum}.
%在一般的阵列综合中，待优化变量为端口数的2倍，与口径的面积成正比。而IDMBSA的优化变量的数与模式数有关，随口径面积的增加较为缓慢。一个粗略的对比图被总结在图 \ref{fig_OptVarNum}中
\subsubsection{Obtain Modal Coefficients by Using MOO}\ 
We obtain $\alpha_{mn}^{i}$ and $\beta_{mn}^{i}$ $(i=x,y)$ by using \textit{the PSO toolbox} (uploaded to \cite{dataset_self}), and define \eqref{eq_corr2} as the cost function. 
The parameter settings for PSO are as follows: the population size is set to 40, with a maximum iteration limit of 100. The initial inertia weight is 0.9, gradually decreasing to 0.4 as the number of iterations increases. 
The cognitive and social learning factors are set to 0.5 and 1.25. The initial positions and velocities of the population are randomly generated in the solution space.
% 在PSO的过程中，种群大小被设置为40，最大迭代次数限制为100次，初始惯性权重为0.9，随着迭代次数增加惯性权重逐渐减少至0.4；认知学习因子设置为0.5，社会学习因子设置 1.25 ，初始种群的位置和速度在解空间中随机生成。

The selection of MOO is not exclusive. This study employed PSO due to its exceptional multi-variable problem adaptability and the author's familiarity with PSO. 
Other MOOs, such as GA \cite{MOO_GA1}, are equally suitable alternatives. 
In fact, directly using the 'interior-point' default optimization method built into the MATLAB \textit{fmincon} function can also achieve optimization (unfortunately, this method is time-consuming and not recommended). 
% 有必要说明，MOO的选择并不唯一，本文中使用PSO的原因一方面是PSO对多变量问题具有较好的适应性，另一方面原因是作者对该方法的调用较为熟悉。其他的多目标优化算法，例如GA[1]也是可行的。实际上，直接使用MATLAB中fmincon函数内置的'interior-point'默认优化方法也可以进行优化（不过不幸的是，这种方法非常耗时，因此不是特别推荐的选择）。

The choice of cost function is not unique and can be selected according to the design requirements. 
Here, we use the Pearson's distance between $|\boldsymbol{f}_\mathrm{tar}(\theta, \phi)|$ and $|\boldsymbol{f}_\mathrm{cal}(\theta, \phi)|$ as the cost function.

%这里我们使用$ |\boldsymbol{f}_\mathrm{tar} (\theta, \phi) | $与 $ |\boldsymbol{f}_\mathrm{cal} (\theta, \phi) | $的皮尔逊距离作为目标函数，
% https://en.wikipedia.org/wiki/Pearson_correlation_coefficient#cite_note-43
% Pearson's distance can be defined from their correlation coefficient as[xx]
%【Fulekar (Ed.), M.H. (2009) Bioinformatics: Applications in Life and Environmental Sciences, Springer (pp. 110) ISBN 1-4020-8879-5】
\setcounter{equation}{11}
\begin{equation}\label{eq_corr2}
obj=1-\rho_{X, Y} 
=1-\frac{\sum_{i=1}^n\left(x_i-\bar{x}\right)\left(y_i-\bar{y}\right)}{\sqrt{\sum_{i=1}^n\left(x_i-\bar{x}\right)^2 \sum_{i=1}^n\left(y_i-\bar{y}\right)^2}}
\end{equation}
where  $ X =|\boldsymbol{f}_\mathrm{tar} (\theta, \phi) |$  and $Y = |\boldsymbol{f}_\mathrm{cal} (\theta, \phi) | $. 
$x_i$ represents an element in $X$, and $\bar{x}$ denotes the average of all elements in $X$, similarly, $y_i$ and $\bar{y}$ follow the same logic. 
$ |\boldsymbol{f}_\mathrm{tar} |$ is known from \eqref{eq_module} and \eqref{eq_module_hv}, and
$ \boldsymbol{f}_\mathrm{cal} $ is calculated from the optimized $\alpha_{mn}^{y}$ and $\beta_{mn}^{y}$.

\begin{figure*} [!htb]% 超长公式
	\setcounter{equation}{10}
	\begin{subequations}\label{eq10}
		\begin{equation}\label{eq10_a}
			\begin{aligned}
				\left|f_x(\theta ,\phi )\right| = \Bigg| \int_{-\frac{b}{2}}^\frac{b}{2}{\int_{-\frac{a}{2}}^\frac{a}{2}{}}&
				\left\{
				\sum_{n=1}^{N}
				\left[
				\left(\sum_{m=1}^{M}\alpha_{mn}^{x}
				\sin \left( \frac{m\pi}{a}(x'+\frac{a}{2}) \right) 
				+ \sum_{m=0}^{M}\beta_{mn}^{x}
				\cos \left( \frac{m\pi}{a}(x'+\frac{a}{2}) \right) 
				\right)
				\sin \left( \frac{n\pi}{b}(y'+\frac{b}{2}) \right) 
				\right]
				\right\}
				\\
				&e^{j(k_x x' + k_y y')}\mathrm{d}x'\mathrm{d}y'  \Bigg|
			\end{aligned}
		\end{equation}
		\begin{equation}\label{eq10_b}
			\begin{aligned}
				|f_y(\theta ,\phi )|= \Bigg|\int_{-\frac{b}{2}}^\frac{b}{2}{\int_{-\frac{a}{2}}^\frac{a}{2}{}} &
				\left\{
				\sum_{m=1}^{M}
				\left[
				\left(\sum_{n=1}^{N}\alpha_{mn}^{y}
				\sin \left( \frac{n\pi}{b}(y'+\frac{b}{2}) \right) 
				+ \sum_{n=0}^{N}\beta_{mn}^{y}
				\cos \left( \frac{n\pi}{b}(y'+\frac{b}{2}) \right)
				\right)
				\sin \left( \frac{m\pi}{a}(x'+\frac{a}{2}) \right)
				\right]
				\right\}
				\\
				& e^{j(k_x x' + k_y y')}\mathrm{d}x'\mathrm{d}y' \Bigg|
			\end{aligned}
		\end{equation}
	\end{subequations}
	\hrulefill
\end{figure*}

\subsection{The IDMBSA Process: A Step-by-Step Guide}
According to \eqref{eq_module}-\eqref{eq_corr2}, IDMBSA can be employed for determining the aperture field for arbitrary phaseless far-field.

The detailed inverse design process is performed as follows:

Step A: Calculate $|f_x|$ and $|f_y|$ using \eqref{eq_module} or \eqref{eq_module_hv} based on the target phaseless far-field radiation pattern. If the far-field polarization requirements are specific, coordinate transformation can be referred to (see \cite{ludwig3}). 
%Step A: 根据 phaseless far-field radiation pattern，使用\eqref{eq_module}或\eqref{eq_module_hv}，计算得到 $ |f_x| $ and $ |f_y| $ 。(若远场的极化要求较为特殊，可参考\cite{ludwig3}实施坐标转换)

Step B: 
Determine certain technical parameters in \eqref{eq10}, such as the number of sampling points for $ \theta $ and $ \phi $. 
For the formulations set up in this paper, it is recommended that M and N take values up to $2a/\lambda$ and $2b/\lambda$, as detailed in Section \ref{sec:sec4B}.
%Step B: 根据口径尺寸a，b，确定\eqref{eq10}中模式数目M、N的取值，an acceptable approach is that the shortest modal wavelength is longer than half of the operating wavelength. (详情见sectionIV B)

Step C: Solve for the unknown coefficients $\alpha_{mn}^{i}$ and $\beta_{mn}^{i}$ in \eqref{eq10} using MOO algorithms. The choice of optimization algorithm and cost function are not unique; we have achieved good results using PSO and \eqref{eq_corr2}.

%Step C：利用MOO求解\eqref{eq10}中的未知系数 $\alpha_{mn}^{i}$ and $\beta_{mn}^{i}$ .( 优化算法选择和设置并不是唯一的，我们使用的是PSO方法和\eqref{eq_corr2}，这取得了较好的效果)

\section{Numerical Simulation}\label{section5}
This section shows numerical experiments using IDMBSA to obtain aperture field that can produce specific $ |\boldsymbol{e}_{\mathrm{tar}}| $. 

\subsection{Design Targets}
\begin{figure}[htb]
	\centering
%	\subfloat[]	{\includegraphics[width=0.33\columnwidth]{./fig/fig_targetEv0.png}}
	\subfloat[]	{\includegraphics[width=0.42\columnwidth]{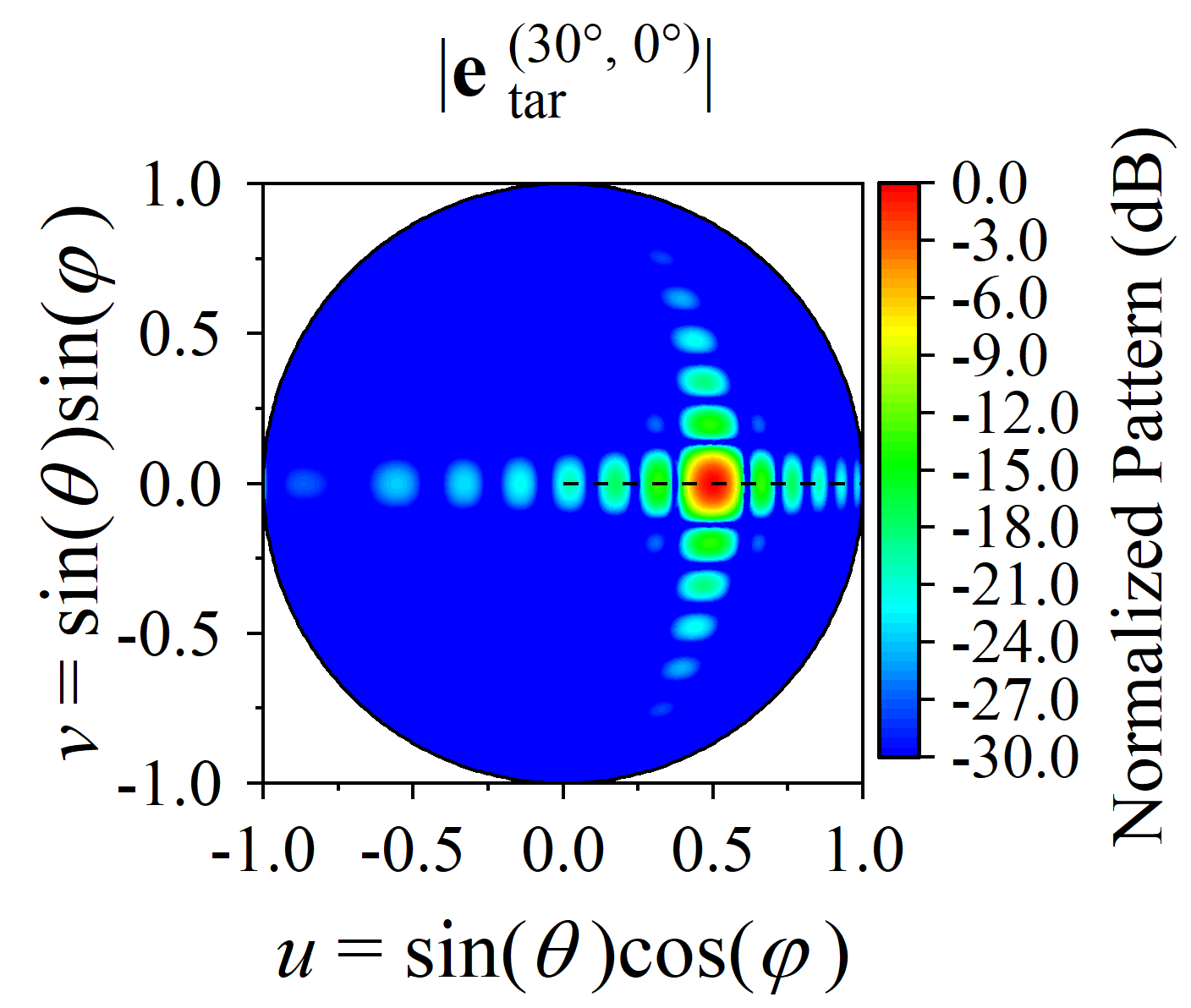}}{  }
	\subfloat[]	{\includegraphics[width=0.42\columnwidth]{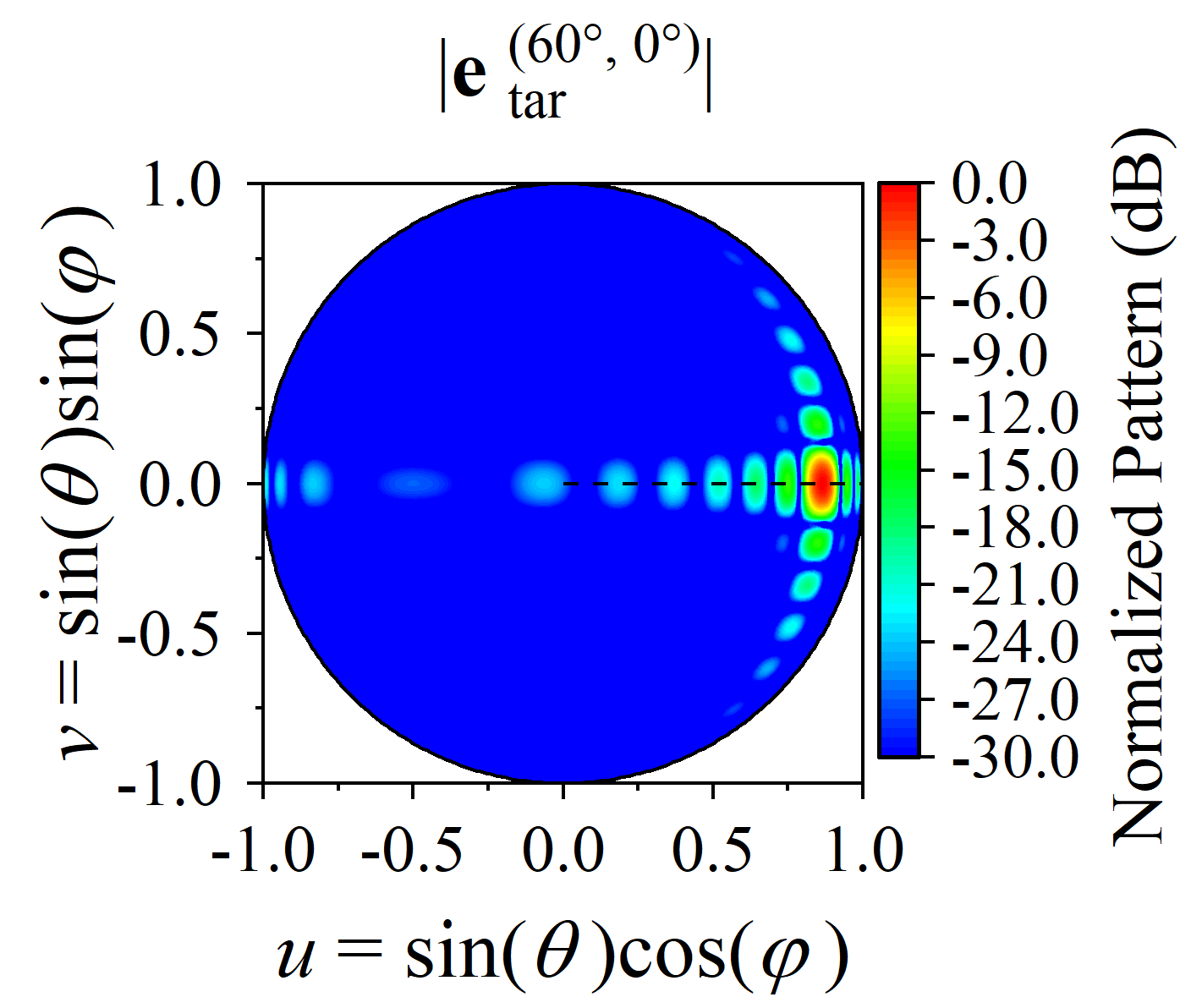}}
	\caption{Target normalized radiation patterns. 
%		(a) $ |\boldsymbol{e}_{\mathrm{tar}}^{(0^\circ,0^\circ)} | $; 
		(a) $ |\boldsymbol{e}_{\mathrm{tar}}^{(30^\circ,0^\circ)} | $;
		(b) $ |\boldsymbol{e}_{\mathrm{tar}}^{(60^\circ,0^\circ)} | $.}
	\label{fig_dipole}
\end{figure}

Here, we employ array factors to generate a set of design targets. 
Consider a 16$\times$16  broadside array located in the $xoy$ plane with uniform amplitude excitation, where the spacing between elements is $\lambda / {2}$. % its three-dimensional radiation pattern is shown in Fig. \ref{fig_dipole}(a), denoted as $|\boldsymbol{e}_{\mathrm{tar}}^{(0^\circ,0^\circ)} |$. 
By directly rotating the coordinate system, we can obtain a set of target patterns with different beam direction angles but the same directivity of 25.2 dBi and side lobe level (SLL) of -13.5 dB. Here, we present two cases, $(30^\circ,0^\circ)$ and $(60^\circ,0^\circ)$, and plot their normalized patterns as shown in Fig. \ref{fig_dipole}. 

Assume that the polarization is only V polarization, represented by \eqref{eq_polar}, as mentioned in case I of section III.
This assumption is not absolute, IDMBSA applies to any polarization distribution. In the following section, we also present a case study on independent design for dual polarization in Section \ref{sec5-c}.
\setcounter{equation}{12}
\begin{subequations}\label{eq_polar}
\begin{align}
	|{e}_{\mathrm{tar},v}^{\text{scan\_angle}}| &= |\boldsymbol{e}_{\mathrm{tar}}^{\text{scan\_angle}}| 
	\\
	{e}_{\mathrm{tar},h}^{\text{scan\_angle}} &= 0
\end{align}
\end{subequations}
Once targets are determined, the first step is to calculate $|f_x|$ and $|f_y|$ by using \eqref{eq_linear}. 
Here, we only consider $|f_y|$, because the calculated $|f_x|$ is approximately 0. Thus, the aperture field has only $E_{\mathrm{ape},y}^{\text{scan\_angle}}$, but no $E_{\mathrm{ape},x}^{\text{scan\_angle}}$.

% 超长公式origin here

\subsection{Numerical Simulation Settings}\label{sec:sec4B}
In the following numerical simulations, the partitioning method for $ \theta $ and $ \phi $ is not absolute and can be adjusted based on the oscillation of the target pattern. Here, $ \theta $ from 0 to $ \pi/2 $ and $ \phi $ from 0 to $ 2\pi $ are divided into 45 and 180 parts, respectively. 

The aperture size $a$ and $b$ are set to $7.5\lambda$, as the same area as the 16$\times$16 array. 
%In conventional array synthesis methods, a 16$\times$16 array typically requires 256 optimization variables (amplitude only) or 512 variables (amplitude and phase). In IDMBSA, only 272 variables are required. 
%在一般阵列综合技术中，16*16的阵列往往需要256个优化变量（仅幅度）或者512个变量（幅度相位同时优化），本方法仅需要272个变量。
Regarding the values of M and N, fewer modes limit design freedom, at times hindering target attainment. 
Excessive modes raise computational load, casusing the field vary dramatically and difficult to achieve. 
%值得特别提到的时，在将IDMBSA应用于阵列综合时，可以根据Nyquist-Shannon定理确定阵列可理论实现的不失真模式的上限。
It is worth noting that when applying IDMBSA to array synthesis, the upper limit of theoretically achievable distortion-free modes in the array can be determined according to the Nyquist-Shannon theorem.
Based on this, the upper limits for M and N should be $a/d_x$ and $b/d_y$, respectively. (
$d_x$ and $d_y$ represent the spacing between array elements.)
%To make the oscillation easy to realize in physics, a trade-off is required. 
% 基于此，M和N的上限取值应该是 $a/d_x$ 和$b/d_y$
%It is recommended that M and N take values up to $2a/\lambda$ and $2b/\lambda$, respectively. 
In this paper, M and N are set to 8, while for the $(60^\circ,0^\circ)$ case they are set to 12. 
%The mode numbers $M$ and $N$ in the following are both set to be $8$. (对于60°的案例，M和Nare set to 12)

%Next step is solving $ E_{\mathrm{ape},y} $ ()from $ |f_y| $ according to \eqref{eq10_b} by using PSO, more precisely, obtaining $\alpha_{mn}^{y}$ and $\beta_{mn}^{y}$ from the $f_y$ calculated above, based on \eqref{eq10b_ana} in Appendix A. 
The next step is solving $ E_{\mathrm{ape},y}^{\text{scan\_angle}} $ (more precisely, obtaining $\alpha_{mn}^{y}$ and $\beta_{mn}^{y}$) from $ |f_y| $ according to \eqref{eq10_b} by PSO. In the iterative optimization process, the integration in \eqref{eq10_b} is accelerated by using its analytical solution \eqref{eq10b_ana}. 
%在MOO的迭代优化中，对于\eqref{eq10_b} 中的积分，使用其解析解\eqref{eq10b_ana} 加速计算

\subsection{Numerical Simulation Results}
\begin{figure}[htb] % fig_aperturefield
	\vspace{-0.8em}
	\centering
	\includegraphics[width=0.8 \columnwidth]{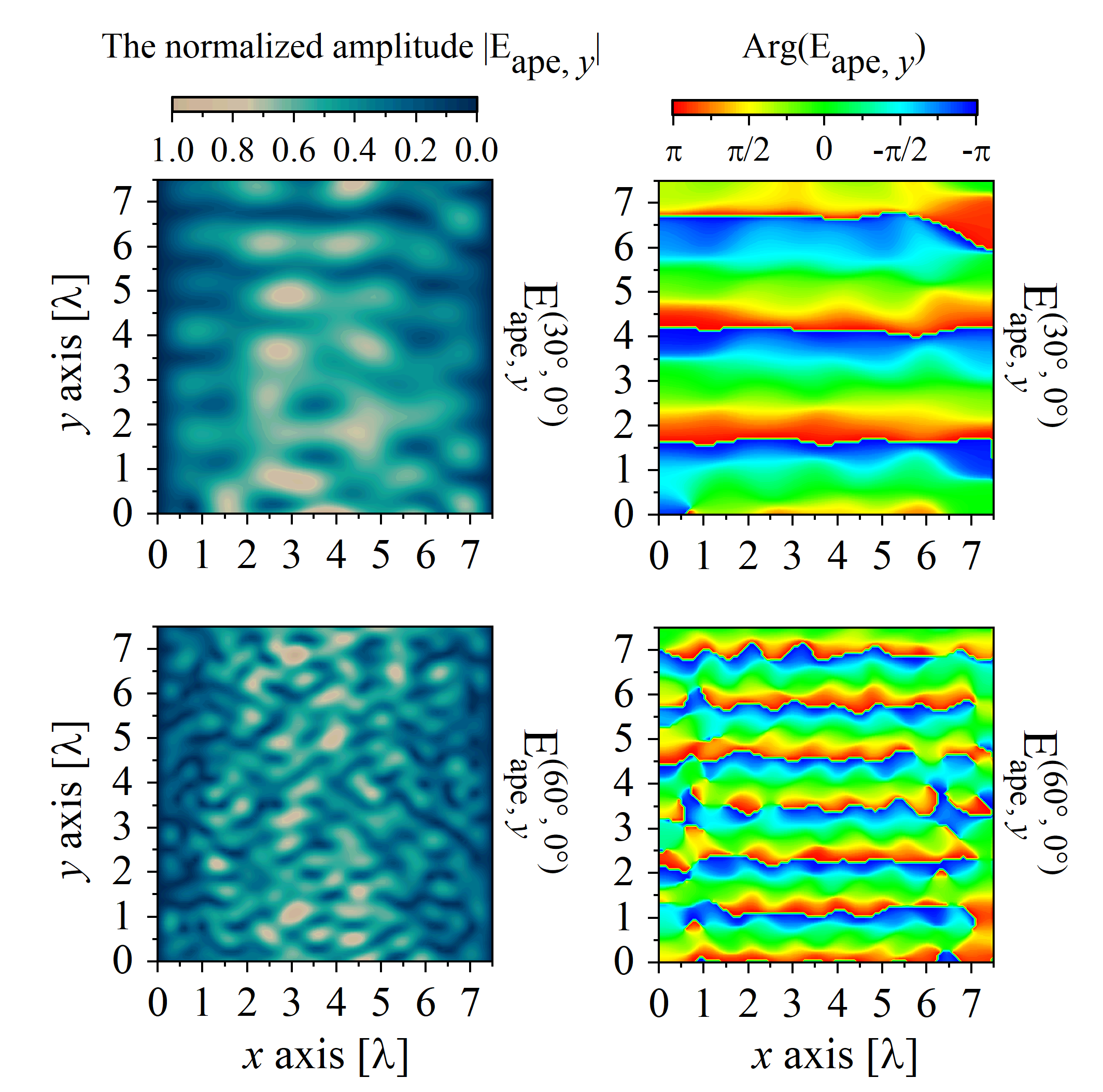}
	\caption{The normalized amplitude distribution and phase distribution of  $E_{\mathrm{ape},y}^{(30^\circ,0^\circ)}$ and $E_{\mathrm{ape},y}^{(60^\circ,0^\circ)}$ . }
	\label{fig_aperturefield}
\end{figure}

\begin{figure}[htb]	% fig_CSTcalfarfield
	\centering
	\subfloat[]	{\includegraphics[width=0.42\columnwidth]{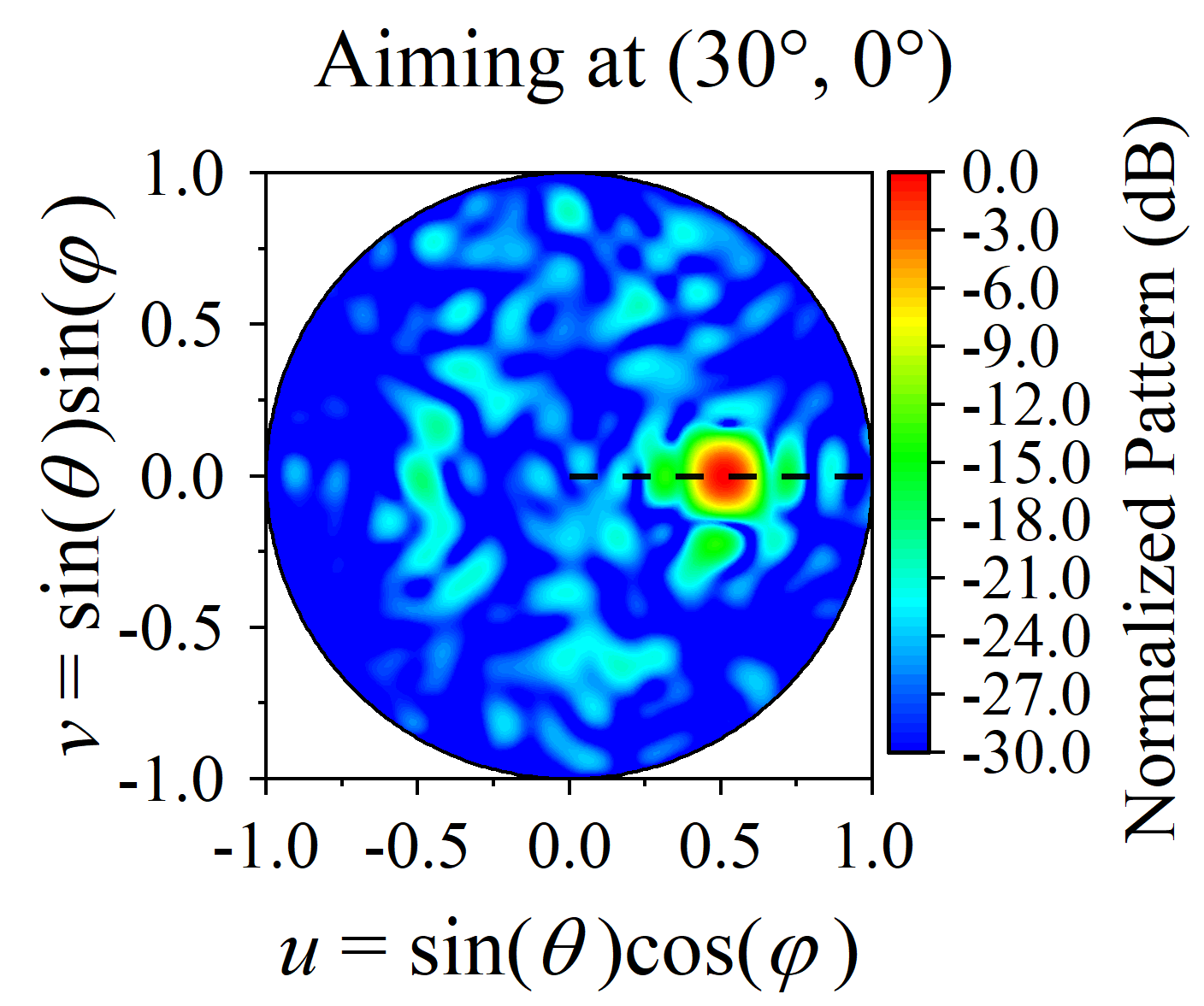}} {  }
	\subfloat[]	{\includegraphics[width=0.42\columnwidth]{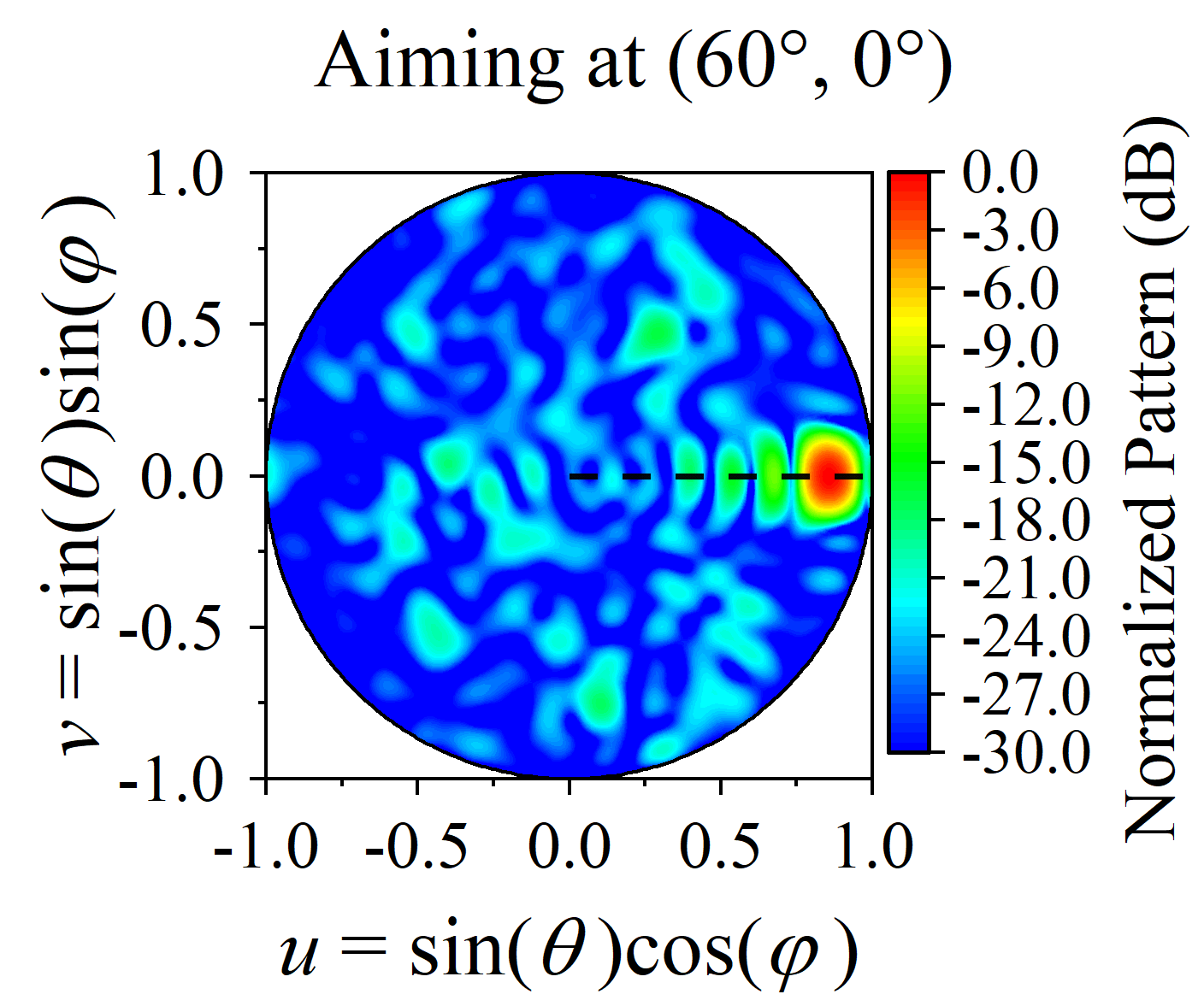}}
	\vspace{-0.8em}
	\\
	\subfloat[]	{\includegraphics[width=0.33\columnwidth]{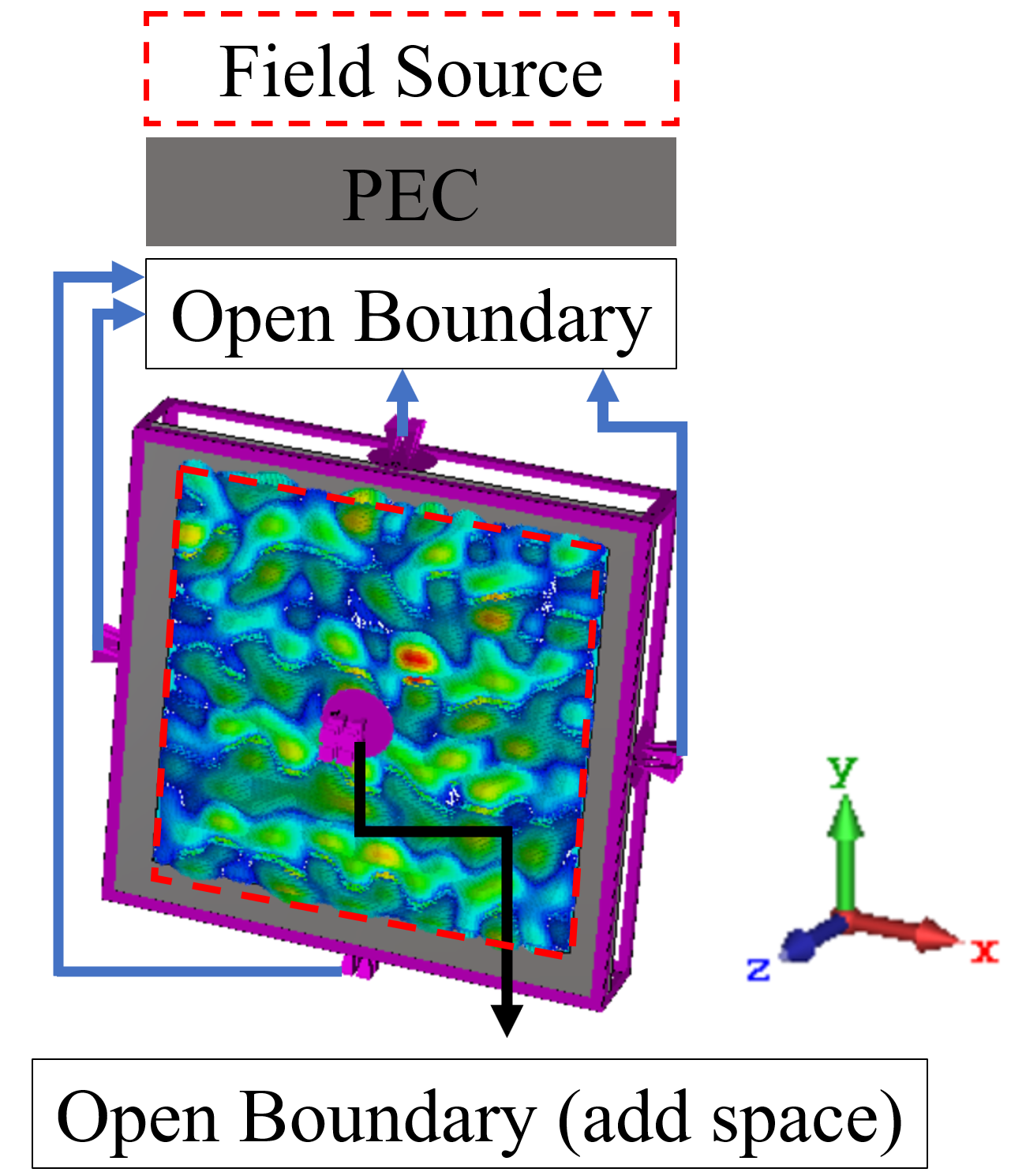}} {  }
	\subfloat[]	{\includegraphics[width=0.63\columnwidth]{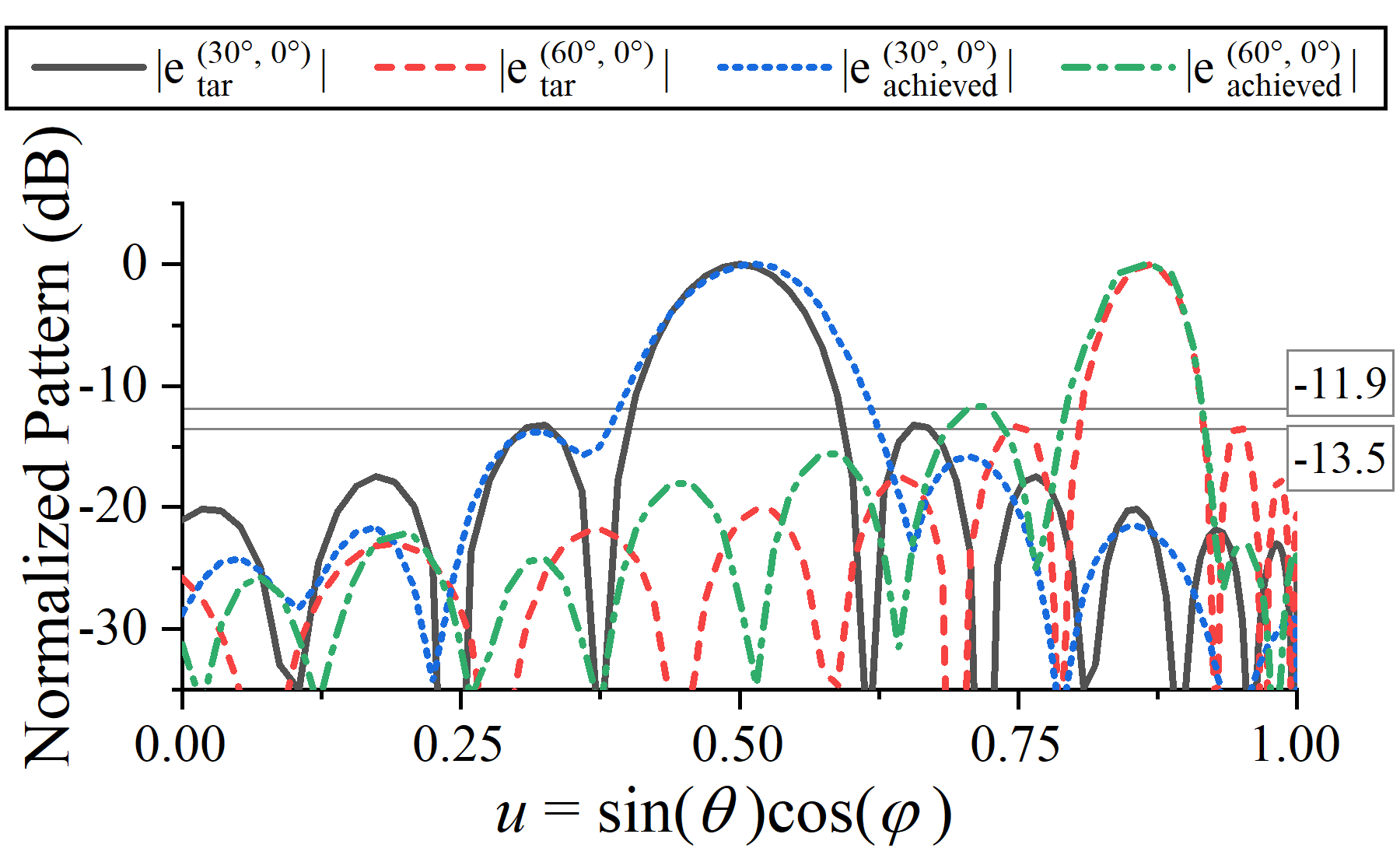 }}
	\caption{ 
		(a) Normalized radiation pattern of $E_{\mathrm{ape},y}^{(30^\circ,0^\circ)}$ ;
		(b) Normalized radiation pattern of $E_{\mathrm{ape},y}^{(60^\circ,0^\circ)}$;
		(c) Simulation settings in \textit{CST Microwave Studio};
		(d) 2D cut-plane: target vs. achieved.}
	\label{fig_CSTcalfarfield}
\end{figure}

The normalized amplitude and phase distribution of $E_{\mathrm{ape},y}^{(30^\circ,0^\circ)}$ and $E_{\mathrm{ape},y}^{(60^\circ,0^\circ)}$  calculated by IDMBSA are shown in Fig. \ref{fig_aperturefield}. 
We performed full-wave simulations using \textit{CST Microwave Studio} to verify the correctness. 
First, constructed $E_{\mathrm{ape},y}$ shown in Fig. \ref{fig_aperturefield} by using scripts to modify files with the extensions ``\textit{nfs}'' and ``\textit{xml}'' from \textit{CST}.
The modified source files were excited in \textit{CST}, and normalized radiation patterns of $E_{\mathrm{ape},y}^{(30^\circ,0^\circ)}$ and $E_{\mathrm{ape},y}^{(60^\circ,0^\circ)}$ are shown in Fig. \ref{fig_CSTcalfarfield}(a, b), with their directivity are 24.2 dBi and 23.9 dBi, respectively. The boundary conditions are set as shown in Fig. \ref{fig_CSTcalfarfield}(c). Additionally, their respective side lobe levels (SLL) are -13.9 dB and -11.5 dB. 
For better visualization, the 2D cut-plane at $ v=0 $ of the dashed lines in Fig. \ref{fig_CSTcalfarfield}(b, c) are summarized in Fig. \ref{fig_CSTcalfarfield}(d). 
The reviewer strongly believes that the number of modes chosen for the modal expansion could be a reason behind the mismatch.
%	为了更清楚的展示，我们展示了pattern 的2D切割图，在v=0处的切割总结在了 Fig. \ref{fig_CSTcalfarfield}(d)中。
The relevant CST files with results have been uploaded to \cite{dataset_self}.

%为了验证由IDMBSA得到的口径场的正确性，我们利用脚本编辑CST软件中的nfs文件和对应的xml文件，将场编辑为由IDMBSA获得的Fig. \ref{fig_aperturefield}。
%在CST软件中将边界条件设置为如图Fig.\ref{fig_CSTcalfarfield}(a)所示，将编辑过后的场源文件导入CST软件并激励，其远场方向图的结果展示在 Fig.\ref{fig_CSTcalfarfield}(b,c)中,相关的CST文件已上传至附件中。


\section{The application of IDMBSA}
In the previous section, we successfully used IDMBSA to design a series of aperture field distributions for achieving specific radiation patterns. However, these aperture fields have not been physically implemented. 
%As mentioned in \cite{inverse_MP_ontheuse}, challenges remain in ensuring the physical readability. 
In this section, we illustrate the physical implementation of the aperture fields through several case studies using microstrip antenna arrays, demonstrate the application of IDMBSA in existing array synthesis and antenna design. Most simulations have been uploaded to \cite{dataset_self}. 
%在上一节中，我们成功使用IDMBSA设计了一系列口径场分布实现特定的远场方向图，但这组口径场尚未被物理实现，因此无法直接应用于工程应用中，正如\cite{inverse_MP_ontheuse} 中提到的“Many challenges remain, including ensuring the resulting  metasurface susceptibilities are physically realizable.”本节我们通过几个案例展示对IDMBSA求解得到的口径场的物理实现，或者说如何将IDMBSA的结果应用于现有的阵列综合中与天线设计中。 
\subsection{Application of IDMBSA in Array Synthesis}
\begin{figure}[htb]	% fig_SingleAntenna
\vspace{-0.8em}
	\centering
	\includegraphics[width=0.7\columnwidth]{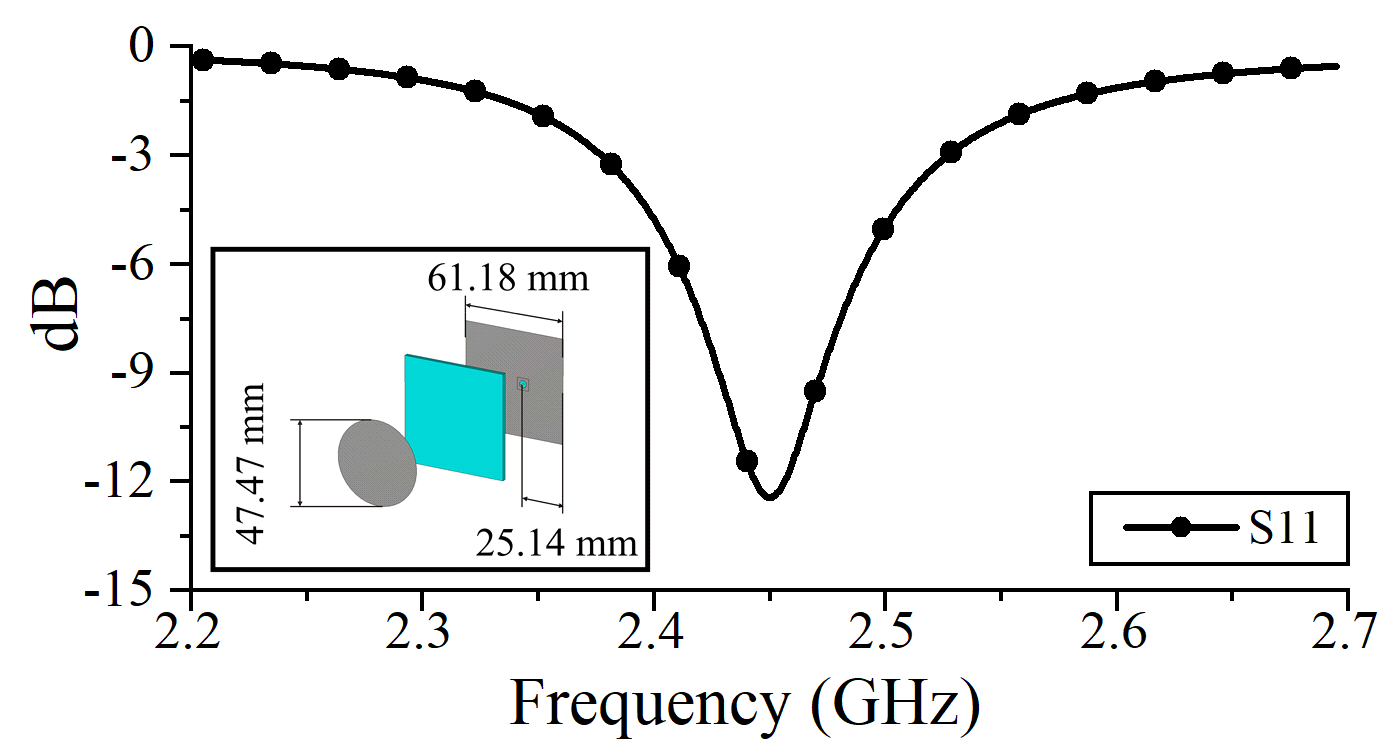}
%	\subfloat[]{\includegraphics[width=0.5\columnwidth]{./fig/fig_SingleAntenna_SingleAndArray.png}}
	\caption{Array element and its S-parameter}
	\label{fig_SingleAntenna}
\end{figure}

In Section IV, the convention of \eqref{eq_polar} leaves the aperture field with only $E_{\mathrm{ape},y}^{\text{scan\_angle}}$. Therefore, a linearly polarized patch antenna is chosen as the array element, and the structure and S-parameters are depicted in Fig. \ref{fig_SingleAntenna}. A planar array of 16$\times$16 elements is arranged, where the inter-element spacing is 61.18mm, half of the wavelength at 2.45GHz. 
%在第IV节中，公式\eqref{eq_polar}的约定使得口径场仅有$E_{\mathrm{ape},y}$分量。因此，在本案例中，选用线极化的贴片天线作为阵列单元。并以$\lambda/2$为间距布置16*16的平面阵列，如图\ref{fig_SingleAntenna}(a)所示，阵元间距为中心频率为2.45GHz波长的一半61.18mm，单元的S参数如图\ref{fig_SingleAntenna}(b)所示。

\begin{figure}[htb]	% fig_ArrayuSynthesis
\vspace{-2em}
	\centering
	\subfloat[]{\includegraphics[width=0.42\columnwidth]{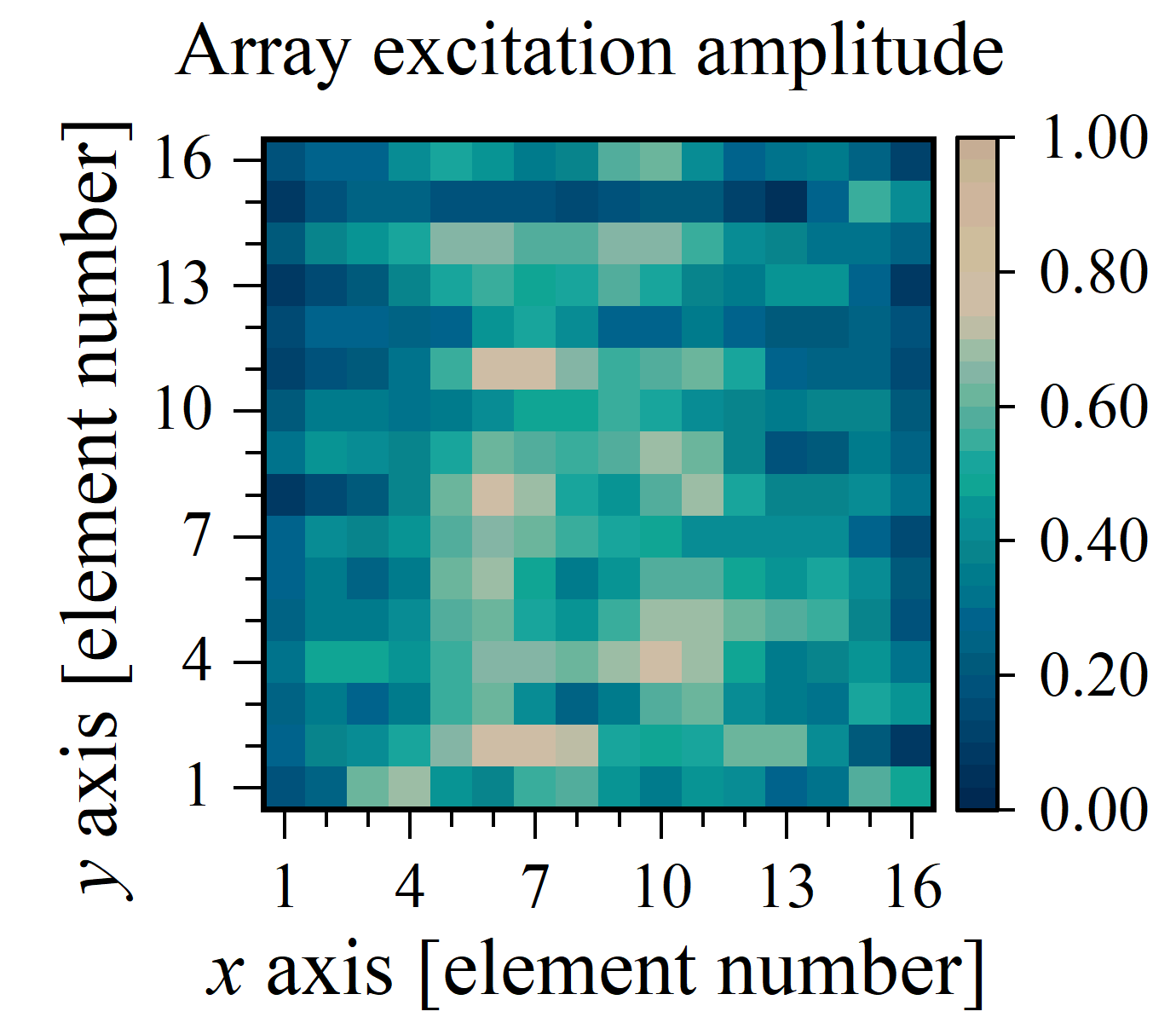}} {  }
	\subfloat[]{\includegraphics[width=0.42\columnwidth]{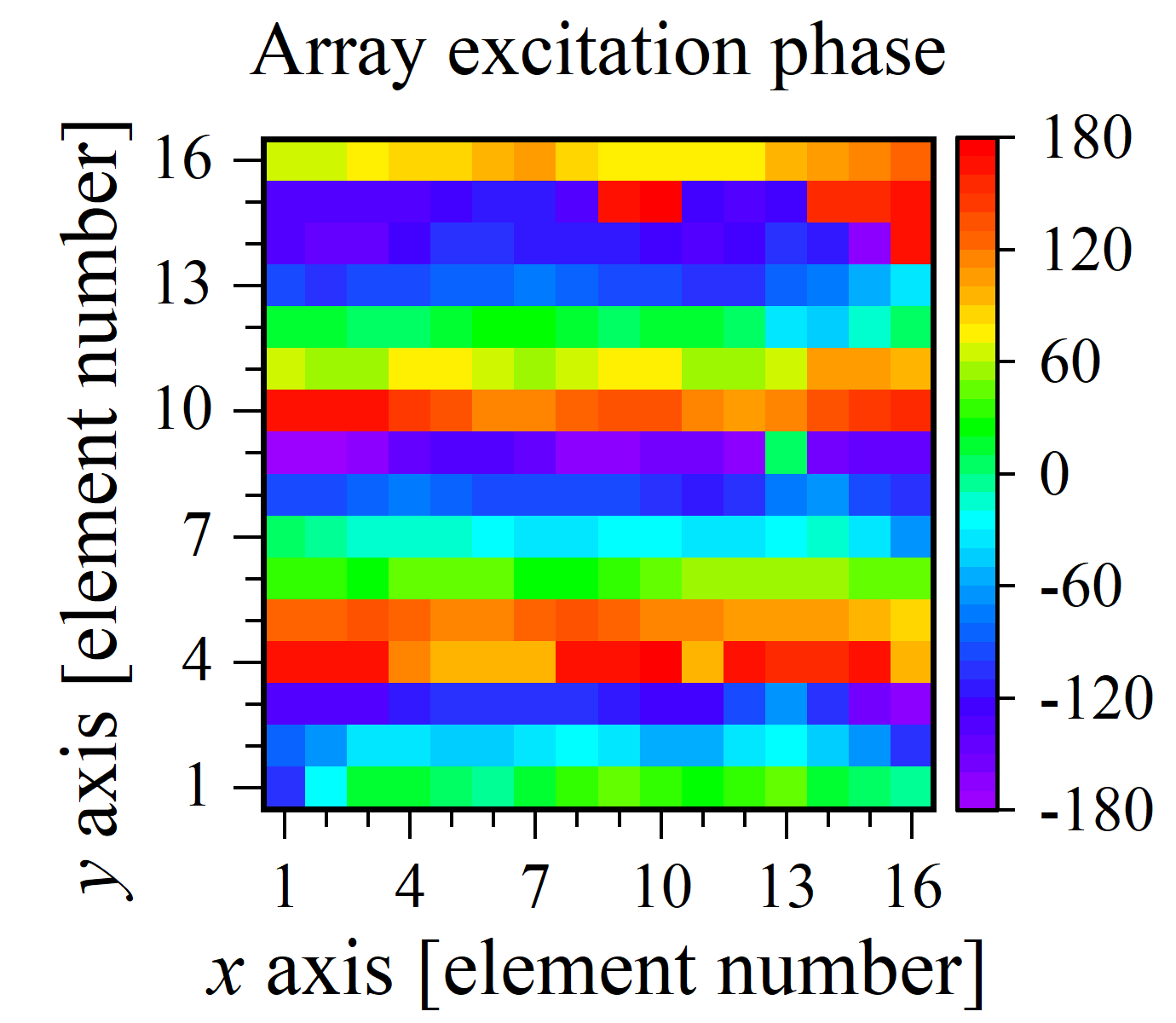}}
	\\
\vspace{-0.8em}
	\subfloat[]{\includegraphics[width=0.42\columnwidth]{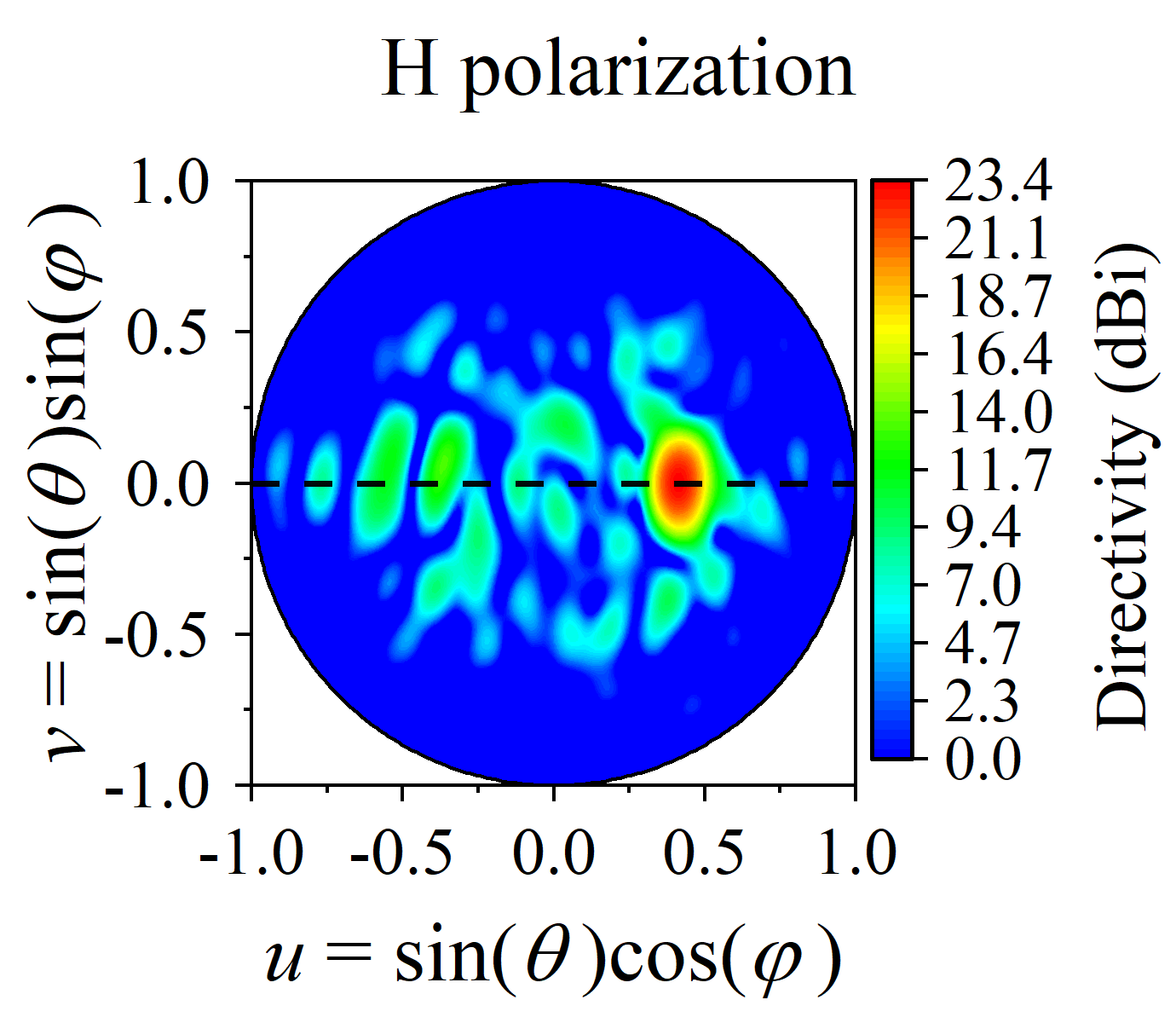}}  {  }
	\subfloat[]{\includegraphics[width=0.42\columnwidth]{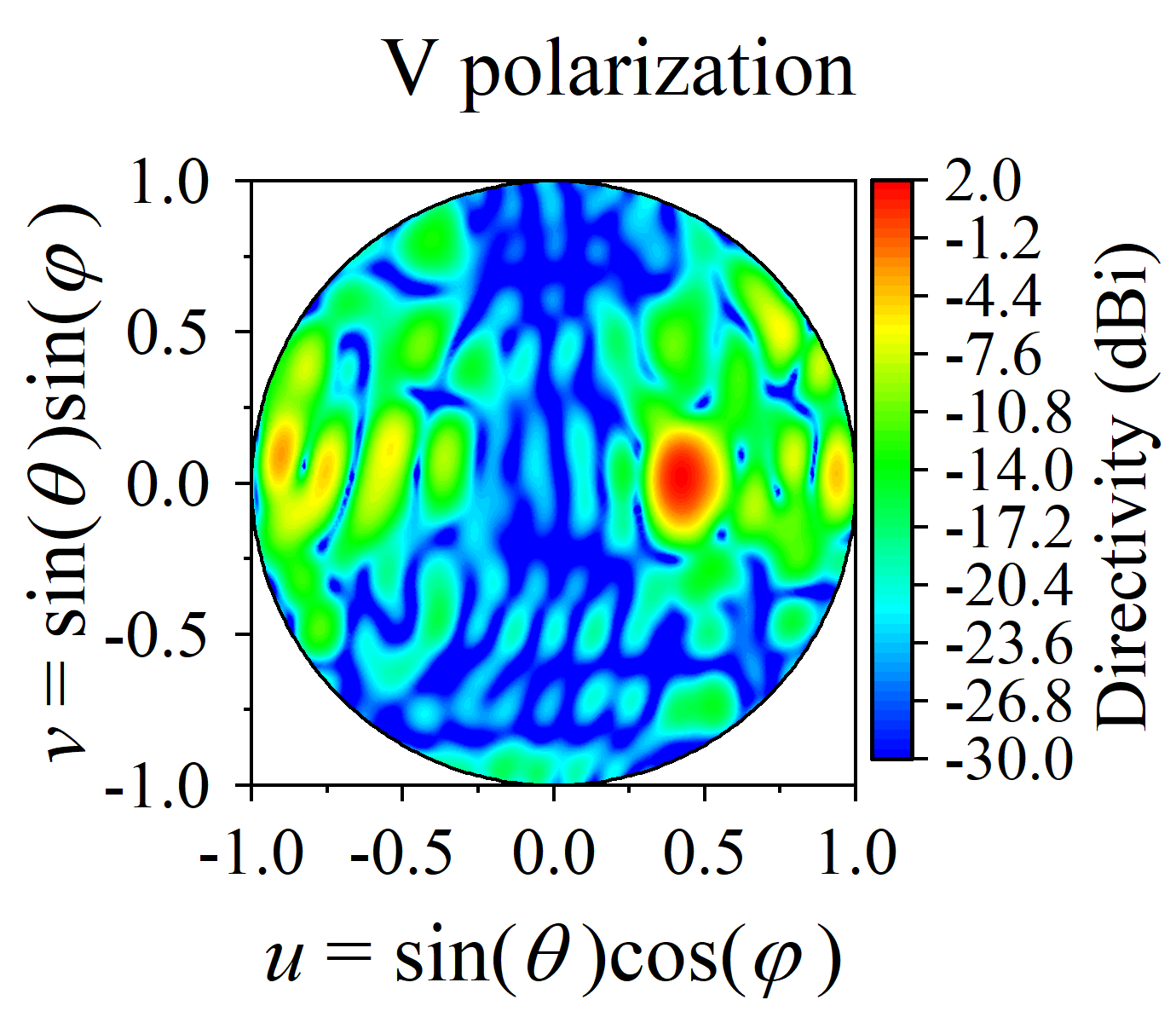}}
	\caption{
		(a) Array excitation amplitude;
		(b) Array excitation phase;
		(c) Array radiation pattern (H polarization);
		(d) Array radiation pattern (V polarization).
	}
	\label{fig_ArrayuSynthesis}
\end{figure}

The feeding amplitudes and phases of each element in the array are provided by the discretized $E_{\mathrm{ape},y}^{(30^\circ,0^\circ)}$. During the discretization process, we use complex electric field at the geometric center of each element as design basis, and the specific values are summarized in Fig. \ref{fig_ArrayuSynthesis}(a, b). 
The different polarization radiation patterns of the array are shown in Fig. \ref{fig_ArrayuSynthesis}(c,d), with a directivity of 23.4 dBi. 
The 2D cut-plane at $ v=0 $ of the dashed lines in Fig. \ref{fig_ArrayuSynthesis}(c) is summarized in Fig. \ref{fig_dualportArray}(f). The simulation results are consistent with the expected pattern shown in Fig. \ref{fig_dipole}(a), indicating that IDMBSA can be applied to array synthesis.  
%阵列中每个单元的馈电幅度和相位由离散后的图\ref{fig_aperturefield}的$E_{\mathrm{ape},y}^{30^\circ}$提供，离散过程中，选用天线单元几何中心对应点的电场的幅度与相位作为依据，具体数值整理在图\ref{fig_ArrayuSynthesis}(a)和(b)中。阵列辐射方向图不同极化分量展示在图\ref{fig_ArrayuSynthesis}(c,d)中。仿真结果与图\ref{fig_dipole}(b)相比符合预期，说明了IDMBSA方法获得的口径场，离散采样后可直接应用于阵列综合中。

\subsection{Application of IDMBSA in Sparse Array Design}
\begin{figure}[htb]	% fig_ArrayuSynthesis_Sparse
	\centering
	\subfloat[]{\includegraphics[width=0.42\columnwidth]{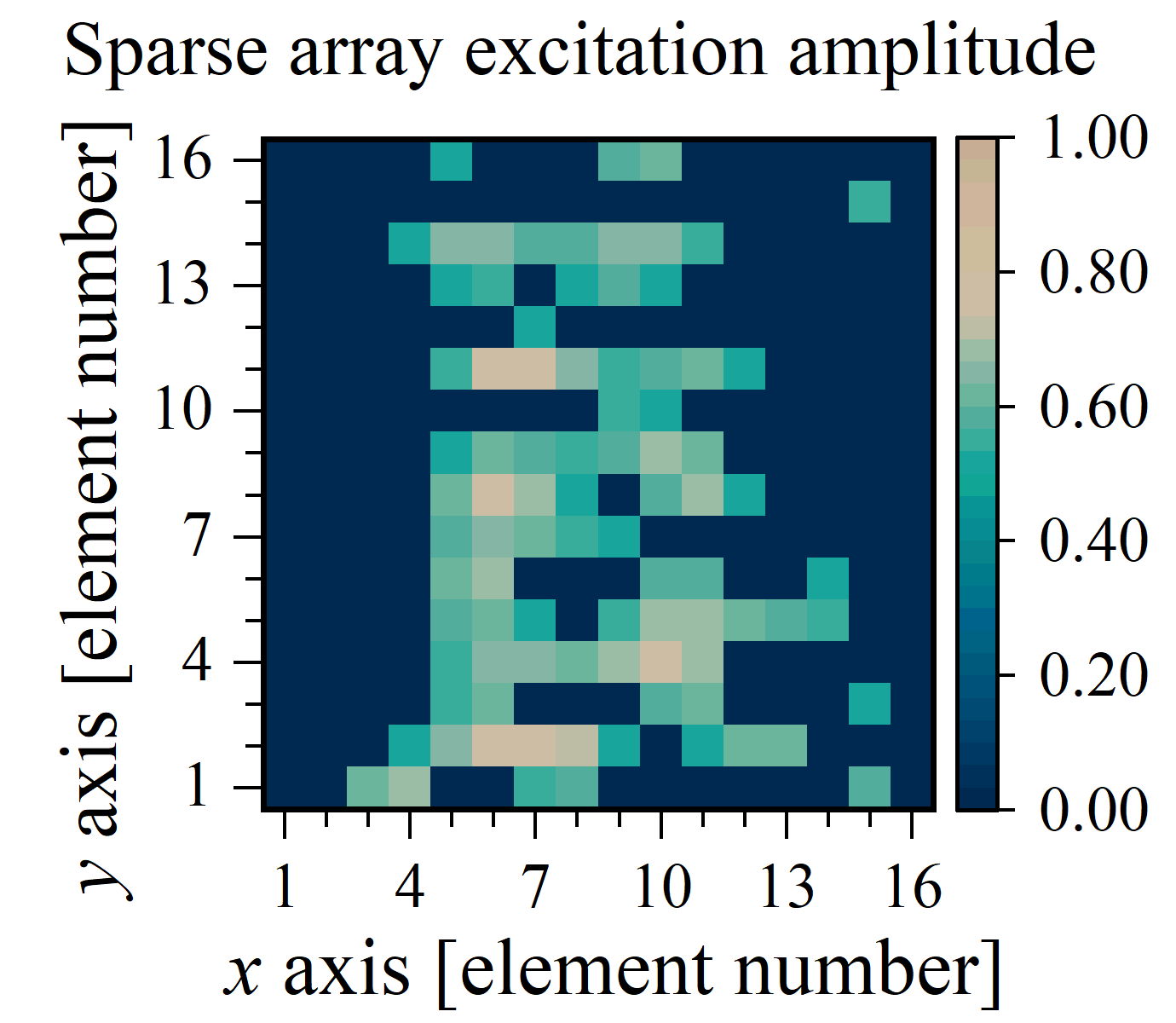}	}{\qquad  }
	\subfloat[]{\includegraphics[width=0.3\columnwidth]{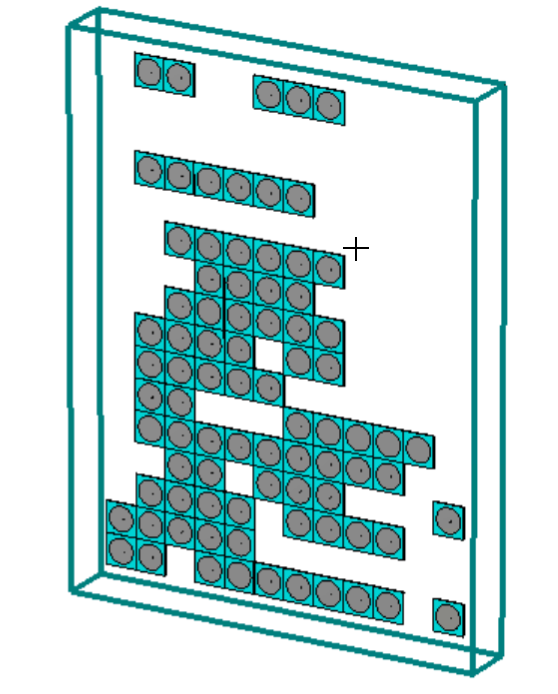}	}
	\\
\vspace{-0.8em}
	\subfloat[]{\includegraphics[width=0.42\columnwidth]{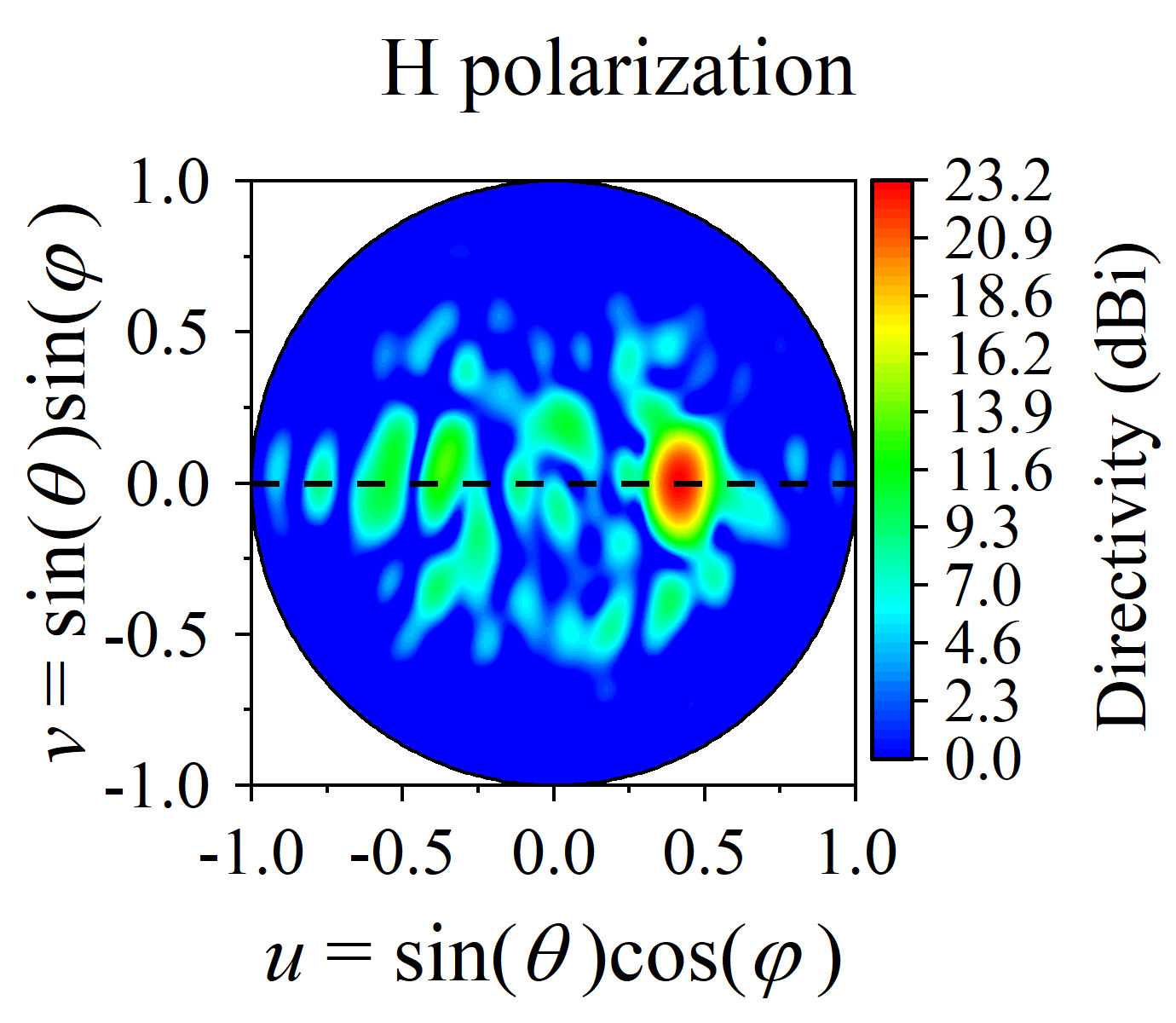}}
	\subfloat[]{\includegraphics[width=0.42\columnwidth]{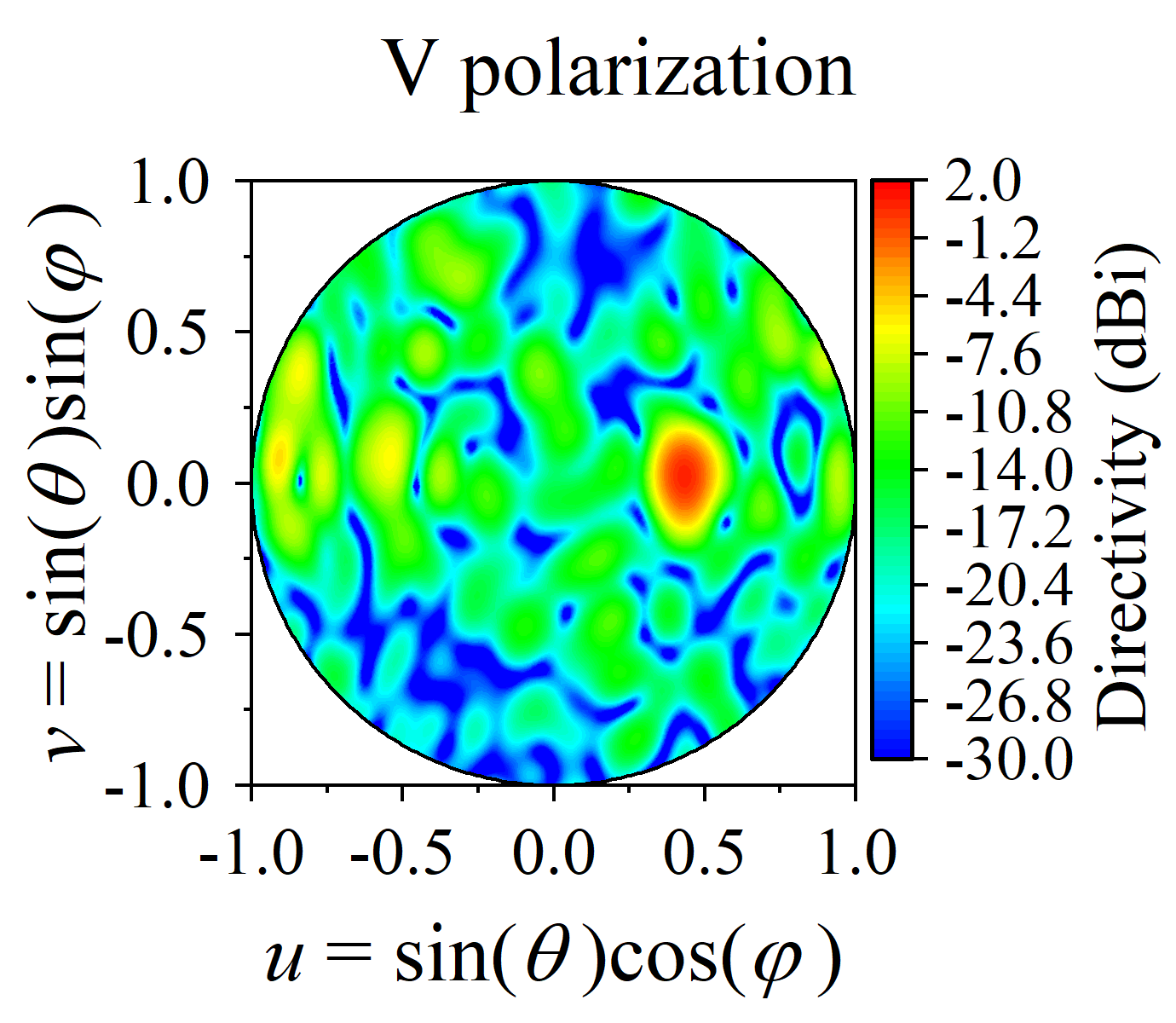}}
	\caption{
		(a) Excitation amplitude of the sparse array;
		(b) Schematic diagram of the sparse array; 
		(c) Radiation pattern of the sparse array (H polarization);
		(d) Radiation pattern of the sparse array (V polarization).
	}
	\label{fig_ArrayuSynthesis_Sparse}
\end{figure}

Observing Fig. \ref{fig_ArrayuSynthesis}(a), it's evident that many array elements have small normalized excitation, allowing us to designate them as dummy elements due to their negligible contributions in the far field. 
Based on numerical experiments, we find that when the total contribution of the low-radiating elements compared to the whole aperture is below 30\%, sparse arrays can effectively replicate the desired performance. 
In this context, we've set the cutoff excitation amplitude at 0.5 (which means that elements with a normalized excitation amplitude less than 0.5 will not be excited), corresponding to the contribution of all low-radiating elements is 25.9\%. 
%Elements with excitation amplitudes greater than 0.5 are activated, and the rest are treated as dummy. 
%阈值的确定应根据the total contribution of the low-radiating elements compared to the whole aperture(简记为$\mathcal{C}$)判断，经过数值实验，对于本案例而言，$\mathcal{C}$小于30\% ，稀布阵都可以较好的复现，当于阈值设置为0.5时，$\mathcal{C}$ is 25.9\%
The specific excitation setting is shown in Fig. \ref{fig_ArrayuSynthesis_Sparse}(a). This approach yields similar radiation pattern. 
Furthermore, by removing the dummy elements, a new sparse array was constructed as shown in Fig. \ref{fig_ArrayuSynthesis_Sparse}(b). It consists of 83 elements and its radiation patterns are shown in Fig. \ref{fig_ArrayuSynthesis_Sparse}(c,d), with a directivity of 23.2 dBi. The 2D cut-plane at $ v=0 $ of the dashed lines in Fig. \ref{fig_ArrayuSynthesis_Sparse}(c) is summarized in Fig. \ref{fig_dualportArray}(f). 

The above case demonstrates that IDMBSA can avoid the optimization of spatial coordinates with traditional sparse array design method\cite{SpareseArray_GA,SpareseArray_GA2}, and can be applied to sparse array design after post-processing (discretization and threshold-based selection). 
%观察图\ref{fig_ArrayuSynthesis}(a)，可以注意到阵列中许多单元的归一化馈电幅度较小，在远场中的贡献可以忽略不计。因此，我们设定一个阈值为0.5，仅激励那些馈电幅度大于0.5的单元。具体的激励情况展示在图\ref{fig_ArrayuSynthesis_Sparse}(a)中，其他单元被设置为哑元并不被激励，可以得到符合预期的远场方向图。
%进一步地，直接移除那些不被激励的哑元，构建了一个新的稀布阵列，如图\ref{fig_ArrayuSynthesis_Sparse}(b)所示，该阵列共包含83个单元。对该稀布阵列进行激励，其远场方向图展示在图\ref{fig_ArrayuSynthesis_Sparse}(c)和(d)中。
%上述案例说明，IDMBSA方法得到的口径场在后处理后（离散化并设置阈值筛选）可应用于稀布阵的设计，避免传统稀布阵设计方法的复杂设计过程[引用（遗传算法稀布阵）]。

\subsection{Independent Design of Dual-Polarization using IDMBSA}\label{sec5-c}
This subsection presents a case of independently designing radiation patterns in both H and V polarization using IDMBSA. 
As we mentioned before, H polarization determines $E_{\mathrm{ape},x}$ and V polarization determines $E_{\mathrm{ape},y}$, so the two polarizations can be designed independently. 
%正如先前提到的，IDMBSA方法适用于多种极化，这里我们展示利用IDMBSA方法对阵列远场方向图双极化分量独立设计的案例。

\begin{figure}[htb]	% fig_ArrayuSynthesis_Sparse
	\centering
%	\subfloat[]{\includegraphics[width=0.5\columnwidth]{./fig/fig_SingleAntenna_dualport.png}}
	\subfloat[]{\includegraphics[width=0.7\columnwidth]{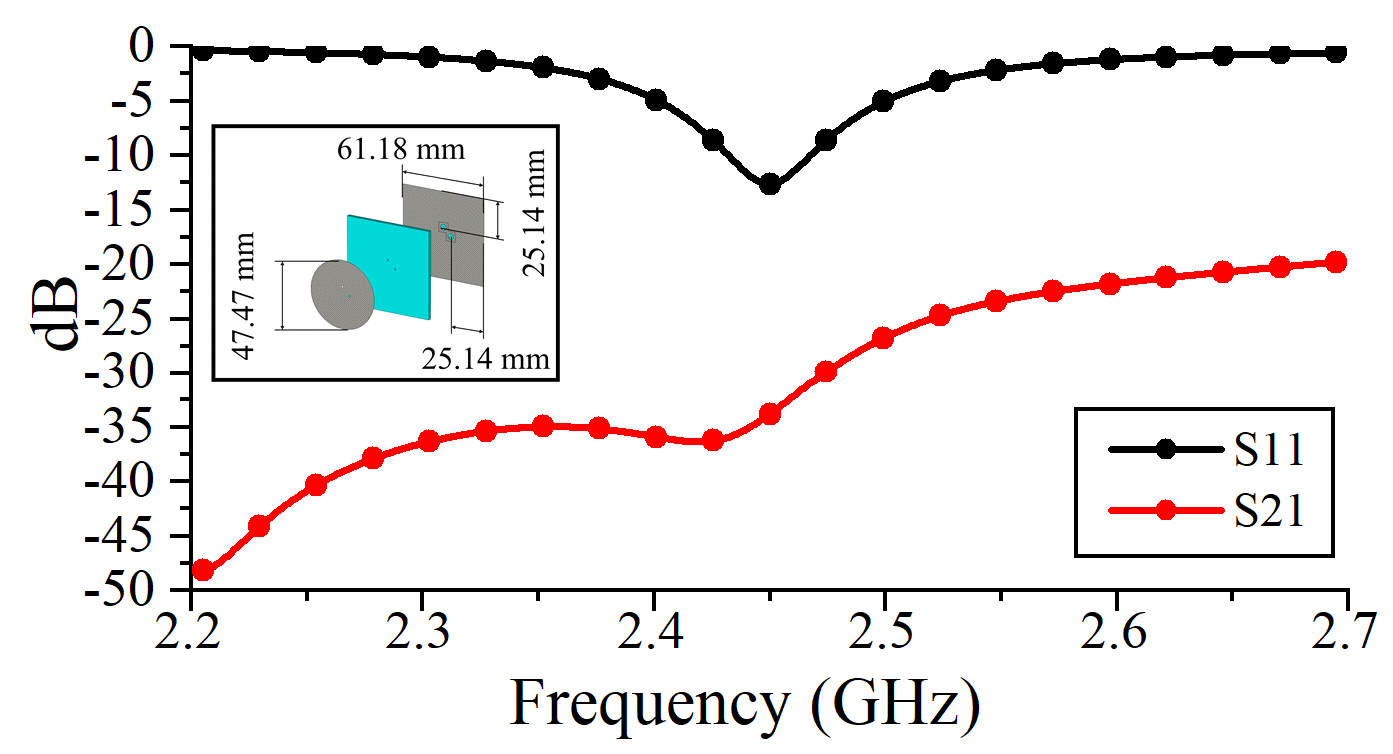}}
\vspace{-0.8em}
	\\
	\subfloat[]{\includegraphics[width=0.4\columnwidth]{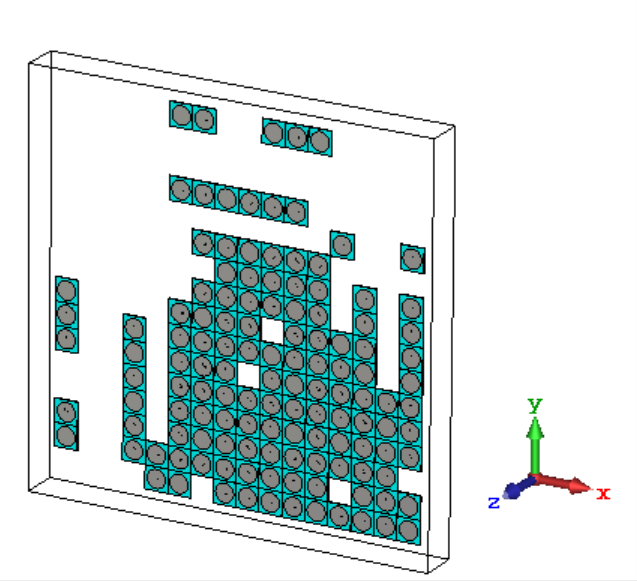}}
	\subfloat[]{\includegraphics[width=0.42\columnwidth]{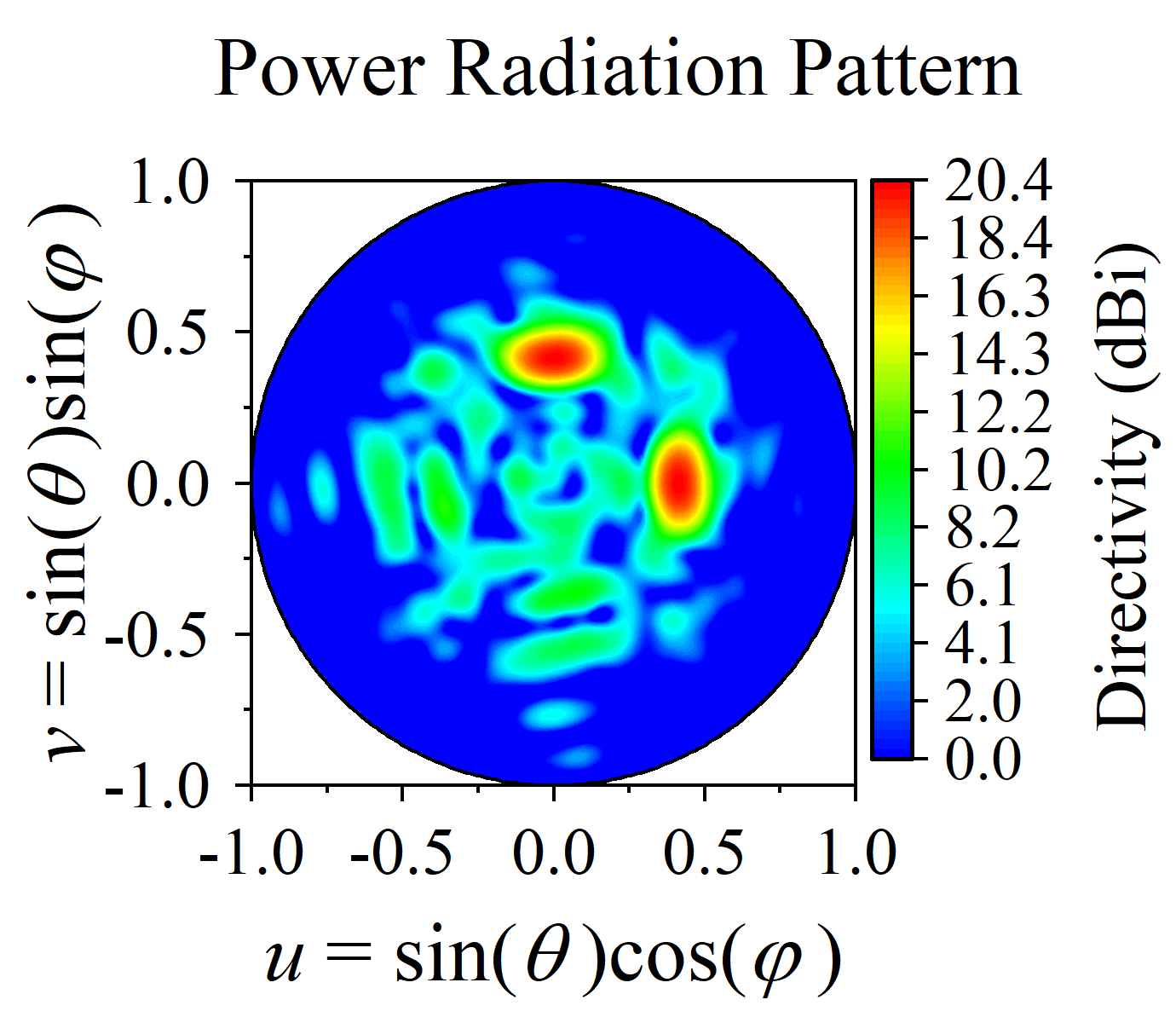}}
	\\
\vspace{-0.8em}
	\subfloat[]{\includegraphics[width=0.42\columnwidth]{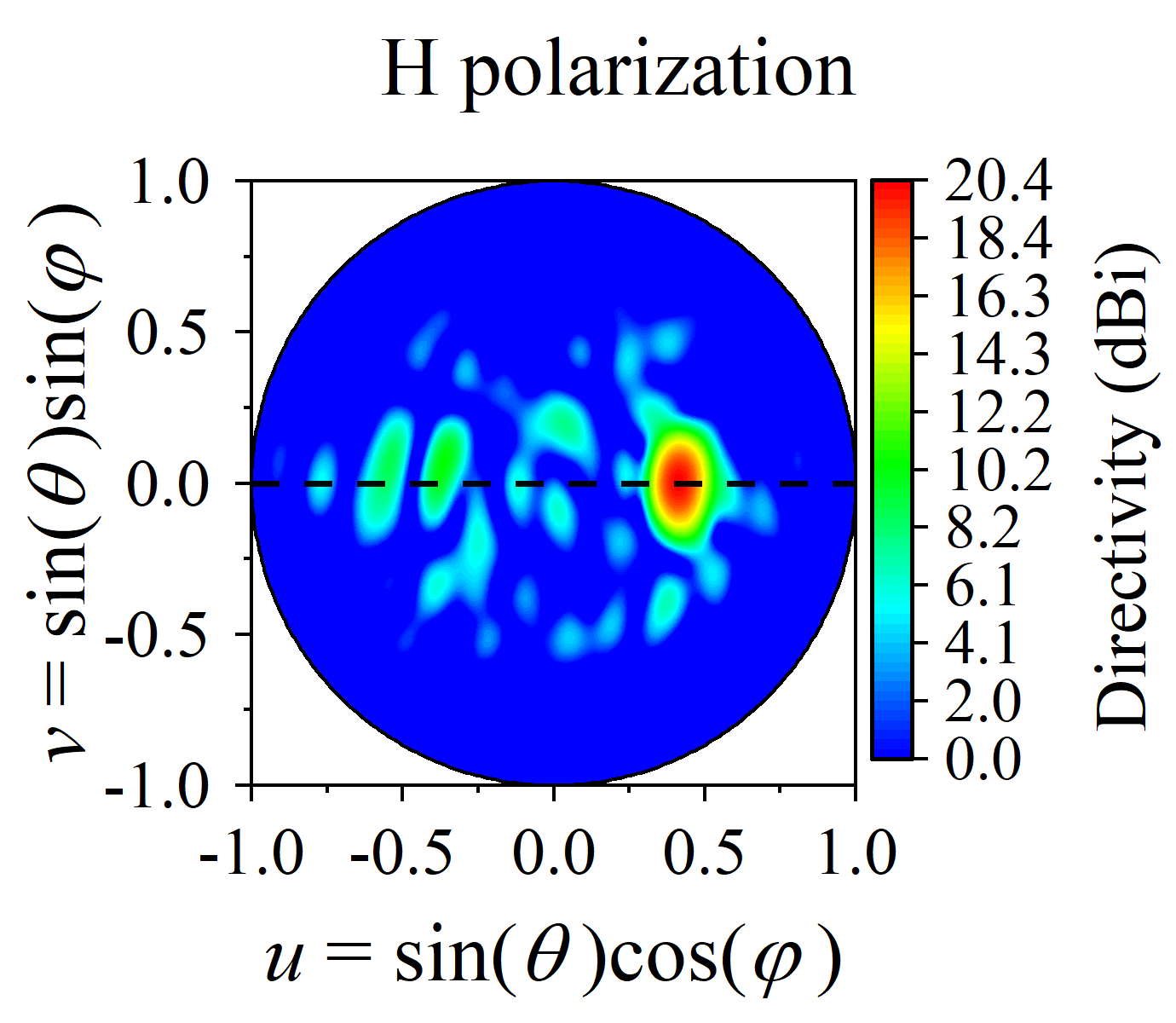}}
	\subfloat[]{\includegraphics[width=0.42\columnwidth]{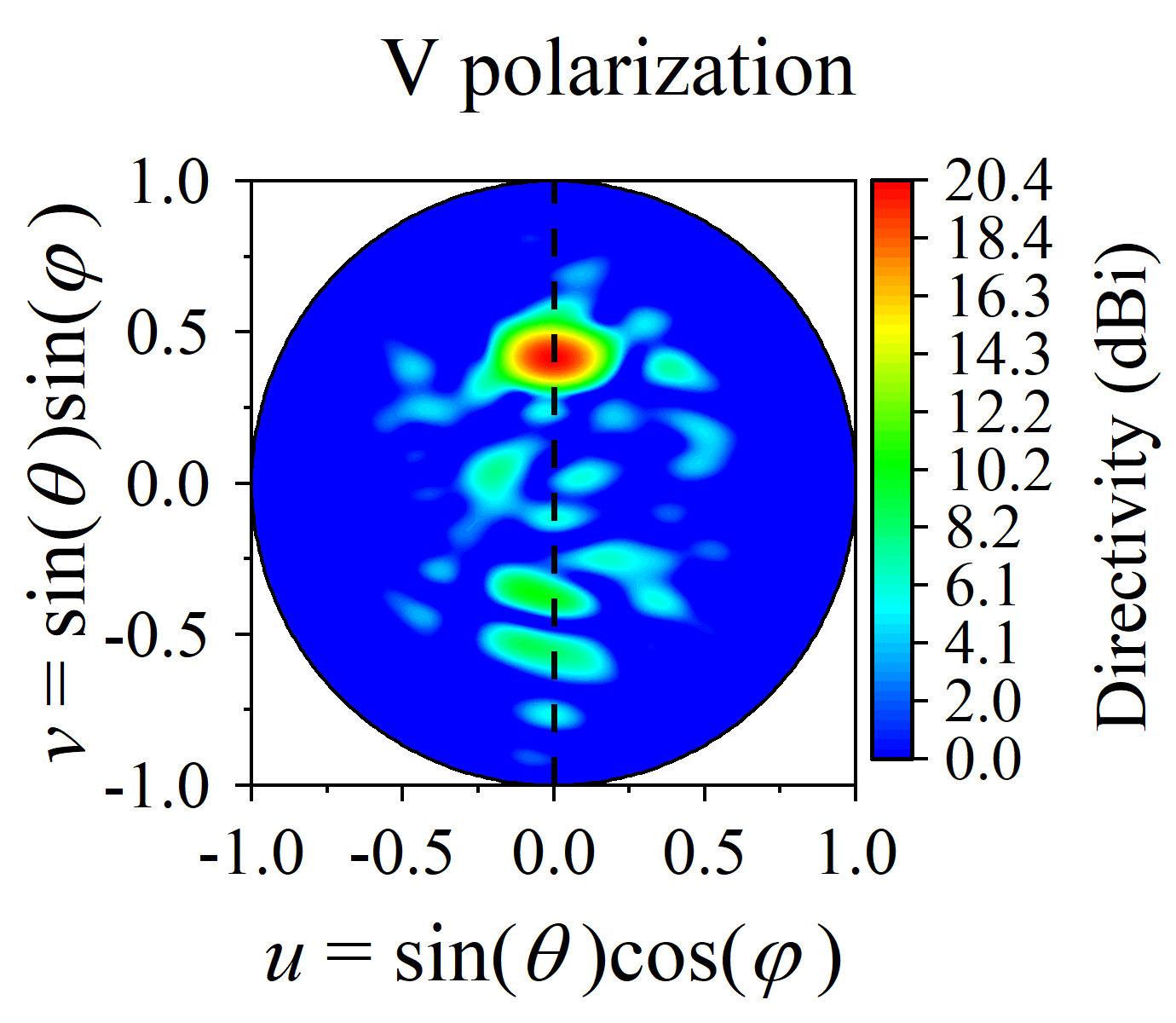}}
	\\
\vspace{-0.8em}
	\subfloat[]{\includegraphics[width=0.8\columnwidth]{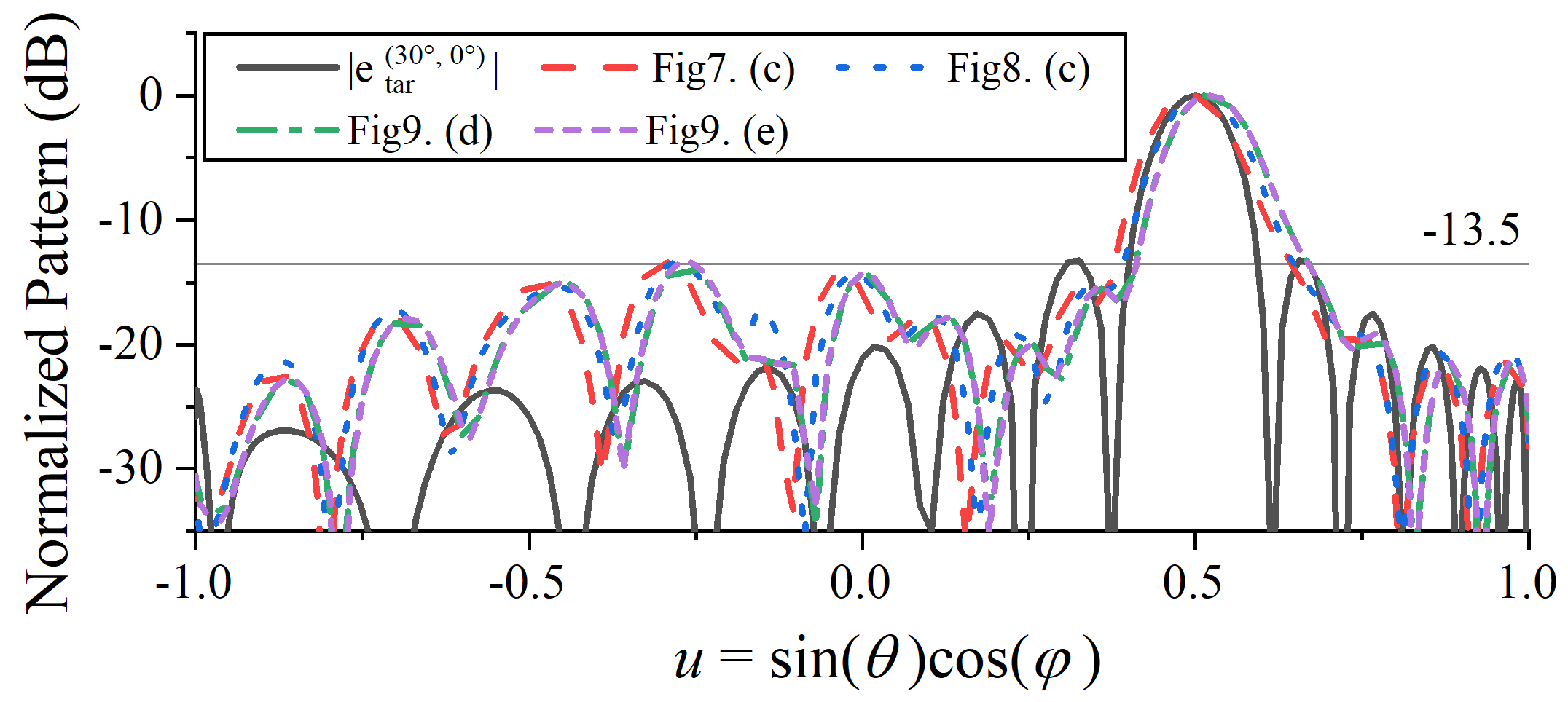}}
	\caption{
		(a) Illustration and S-parameters of the array element
		(b) Schematic diagram of the dual-polarized sparse array;
		(c) Power radiation pattern;
		(d) Radiation pattern (H polarization); 
		(e) Radiation pattern (V polarization);
		(f) 2D cut-plane at $ v=0 $: target vs. achieved. 
	}
	\label{fig_dualportArray}
\end{figure}

In this case, both H and V polarizations are configured to form single beams with different directional angles, as shown in \eqref{eq_example_dual_polar}.
%实际应用中，两个极化的方向图可以独立设定。在本案例中，H极化和V极化都是指向30°的单波束，如公式\eqref{eq_example_dual_polar}所示。
%$E_{\mathrm{ape},x}^{(0^\circ,30^\circ)}$ and $E_{\mathrm{ape},y}^{(30^\circ,0^\circ)}$ can be directly combined. 
%在这种约束下，H极化决定 $E_{\mathrm{ape},x}$，V极化决定 $E_{\mathrm{ape},y}$，考虑到$E_{\mathrm{ape},x}$与$E_{\mathrm{ape},y}$属不同分量，可直接融合，离散化后的结果同理可以直接叠加。
\setcounter{equation}{13}
\begin{subequations}\label{eq_example_dual_polar}
	\begin{align}
		|{e}_{\mathrm{design},v}| &= |\boldsymbol{e}_{\mathrm{tar}}^{(30^\circ,0^\circ)}|
		\\
		|{e}_{\mathrm{design},h} |&=  |\boldsymbol{e}_{\mathrm{tar}}^{(0^\circ,30^\circ)} |
	\end{align}
\end{subequations}
The post-processed  $E_{\mathrm{ape},y}^{(30^\circ,0^\circ)}$ field is consistent with Fig. \ref{fig_ArrayuSynthesis_Sparse}(a), and the post-processed $E_{\mathrm{ape},x}^{(0^\circ,30^\circ)}$ field corresponds to a 90° rotation of Fig. \ref{fig_ArrayuSynthesis_Sparse}(a).
%后处理后的$E_{\mathrm{ape},y}$与图\ref{fig_ArrayuSynthesis_Sparse}(a)一致，后处理后的$E_{\mathrm{ape},x}$为图\ref{fig_ArrayuSynthesis_Sparse}(a)中的矩阵旋转90°。
We verify this using both planar array and sparse array, and get similar results. For brevity, we present the results of the sparse array here.
%我们使用平面阵列和稀疏阵列均实现了设计目标，二者结果接近，出于篇幅原因此处展示稀疏阵列的结果。
The array element is a modified patch antenna with dual-port feeding. The ports are adjusted to ensure the physical installation of SMA connectors, and its structure and S-parameters are shown in Fig. \ref{fig_dualportArray}(a). The array is illustrated in Fig. \ref{fig_dualportArray}(b).
%阵列单元选用微调后的贴片天线，端口调整为双端口馈电，且确保双端口的SMA接头可物理安装，具体结构如图\ref{fig_dualportArray}(a)所示，该单元双端口的隔离度展示在图\ref{fig_dualportArray}(b)中，阵列结构如图\ref{fig_dualportArray}(c)所示。
The array consists of 136 elements and 272 ports, but only 166 ports are utilized. Its radiation patterns are shown in Fig. \ref{fig_dualportArray}(c,d,e). Additionally, their directivities are 20.4 dBi. The 2D cut-plane at $ v=0 $ of the dashed lines for both polarizations are summarized in Fig. \ref{fig_dualportArray}(f). 
The observed mismatch in Fig. \ref{fig_dualportArray}(f) between the results and the objectives on sidelobes and nulls may be attributed to an increased number of selected modes, thus hindering the accurate replication of the aperture field by the array. 
%对于Fig. \ref{fig_dualportArray}(f) 中结果与目标在旁瓣和nulls上的不匹配，我们认为这可能是模式数量选择的较多使得口径场的无法精确的由阵列复现导致的。
%The case demonstrates the application of IDMBSA in the independent design of dual-polarization.
%阵列共136个单元,其中仅166个端口被使用。图\ref{fig_dualportArray}(c)所示阵列的远场辐射方向图展示在图\ref{fig_dualportArray}(d,e,f)中。不难发现我们实现了对两个极化方向图的独立设计。
%案例的成功说明了IDMBSA在对于实际工程应用具有很强的指导意义。

\subsection{Other Potential Applications}
\subsubsection{Flat Topped Beam}
\begin{figure}[htb]	
	\centering
	\subfloat[]{\includegraphics[width=0.42\columnwidth]{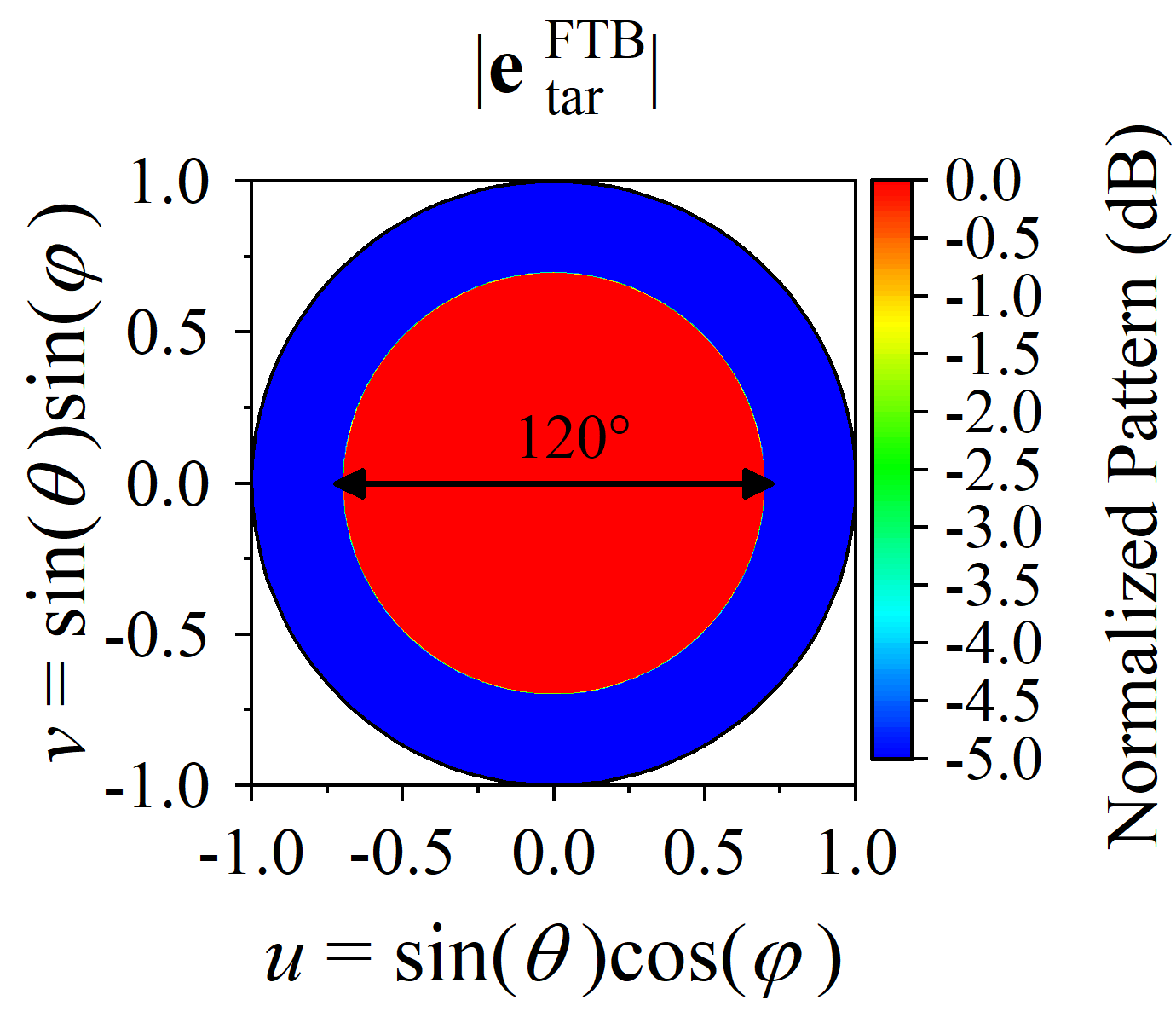}}{  }
	\subfloat[]{\includegraphics[width=0.42\columnwidth]{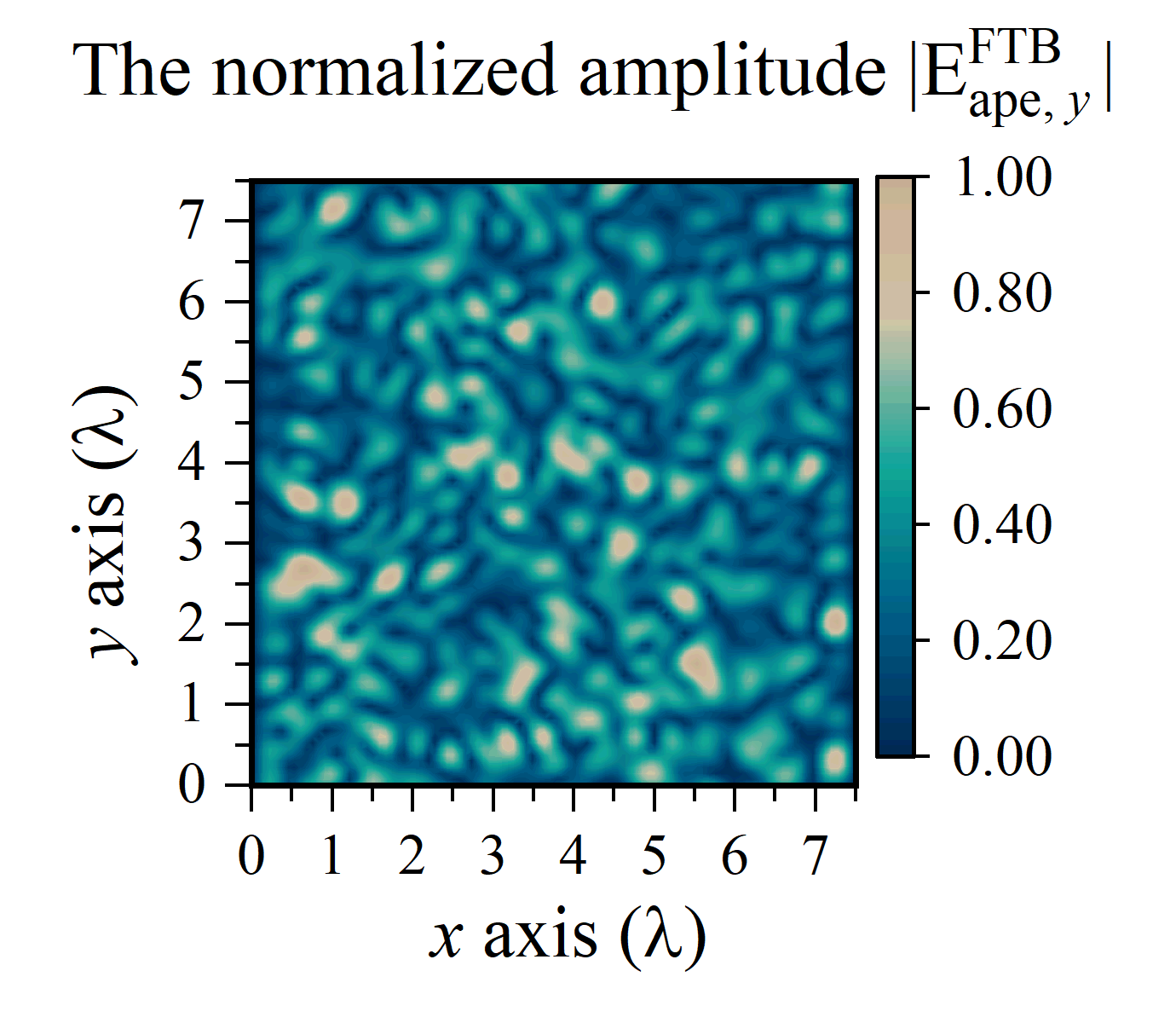}}
	\vspace{-0.8em}
	\\
	\subfloat[]{\includegraphics[width=0.42\columnwidth]{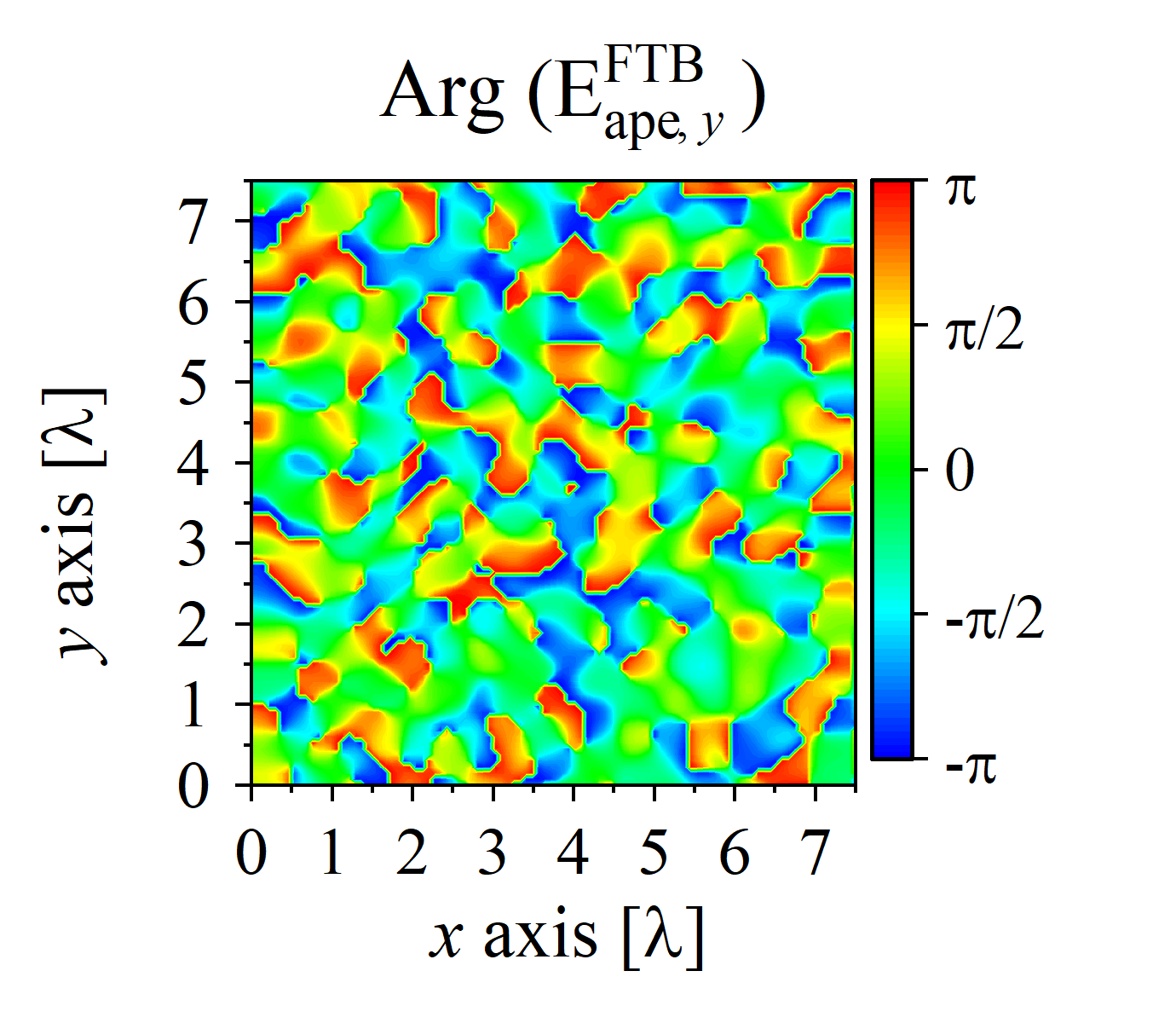}}
	\subfloat[]{\includegraphics[width=0.42\columnwidth]{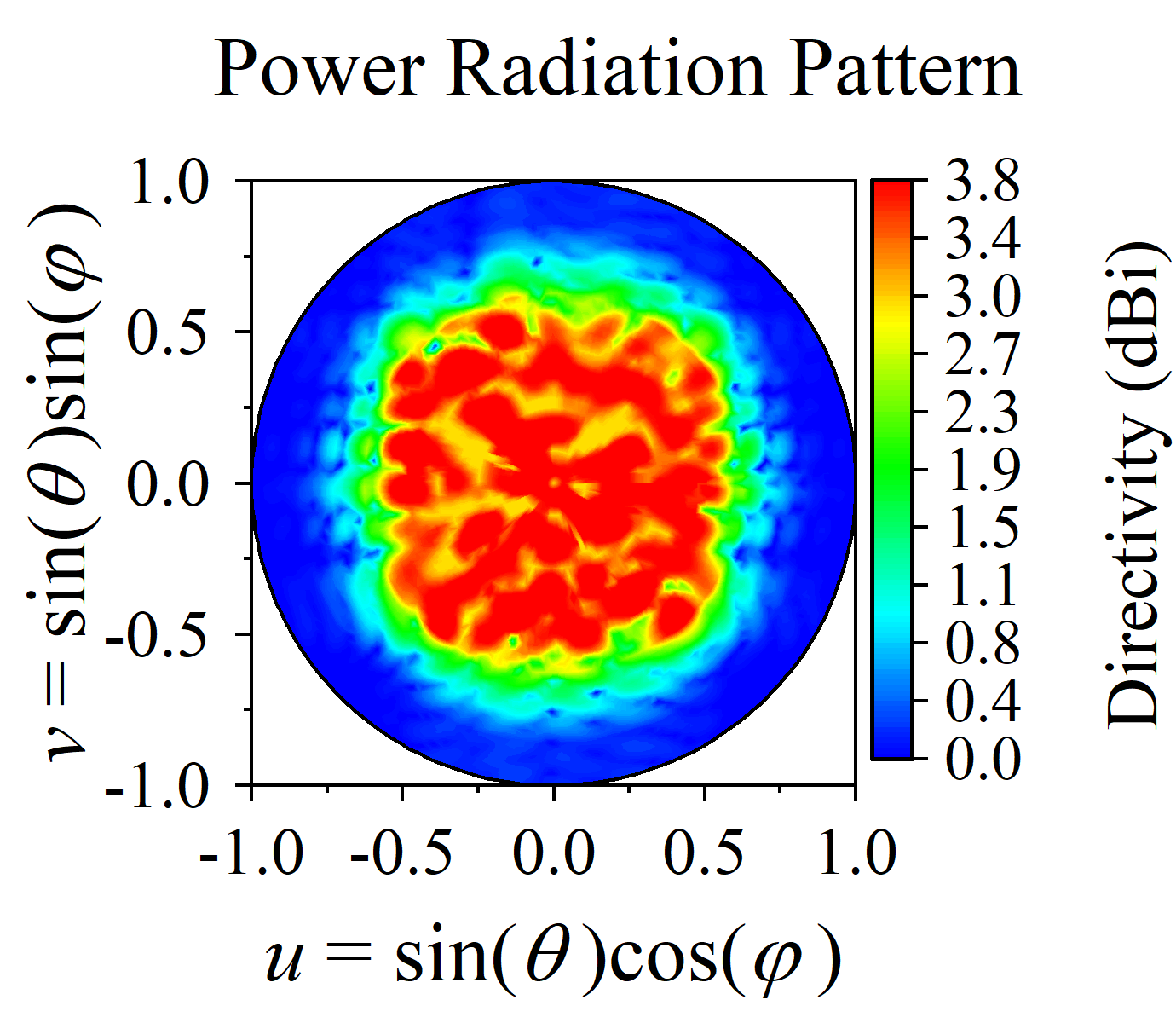}}
	\caption{
		(a) Target FTB pattern;
		(b) $\left| E_{\mathrm{ape},y}^{\mathrm{FTB}} \right|$;
		(c) Arg $\left( E_{\mathrm{ape},y}^{\mathrm{FTB}} \right) $;  
		(d) Power radiation pattern of $ E_{\mathrm{ape},y}^{\mathrm{FTB}}$. 
	}
	\label{fig_ftb}
\end{figure}

The IDMBSA is also suitable for designing special patterns, and it only requires modifying the target pattern. 
Here, we demonstrate a case using the flat topped beam (FTB) pattern as shown in Fig.\ref{fig_ftb}(a) as the target. The polarization constraint remains \eqref{eq_polar}.  
The aperture field obtained using IDMBSA is denoted as $E_{\mathrm{ape},y}^{\mathrm{FTB}}$ as shown in Fig. \ref{fig_ftb}(b, c). Its radiation pattern is shown in Fig. \ref{fig_ftb}(d). 
% IDMBSA方法同样适用于特殊波束的设计，只需修改目标方向图即可。在这里，我们以图x所示的平顶波束为目标，展示一个案例。

\subsubsection{Low Side Lobe Level}
\begin{figure}[htb]	
	\centering
	\subfloat[]{\includegraphics[width=0.42\columnwidth]{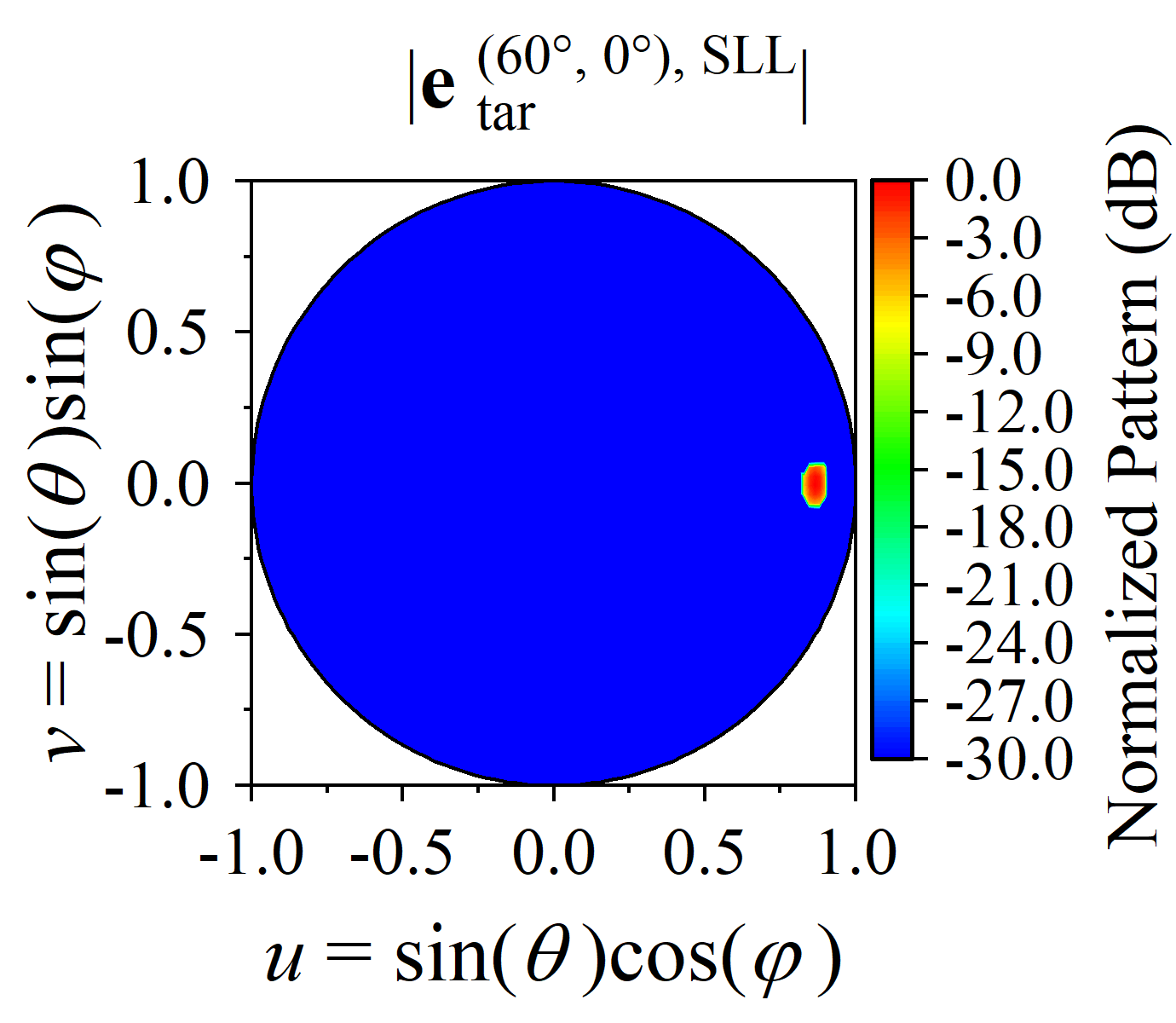}}{  }
	\subfloat[]{\includegraphics[width=0.42\columnwidth]{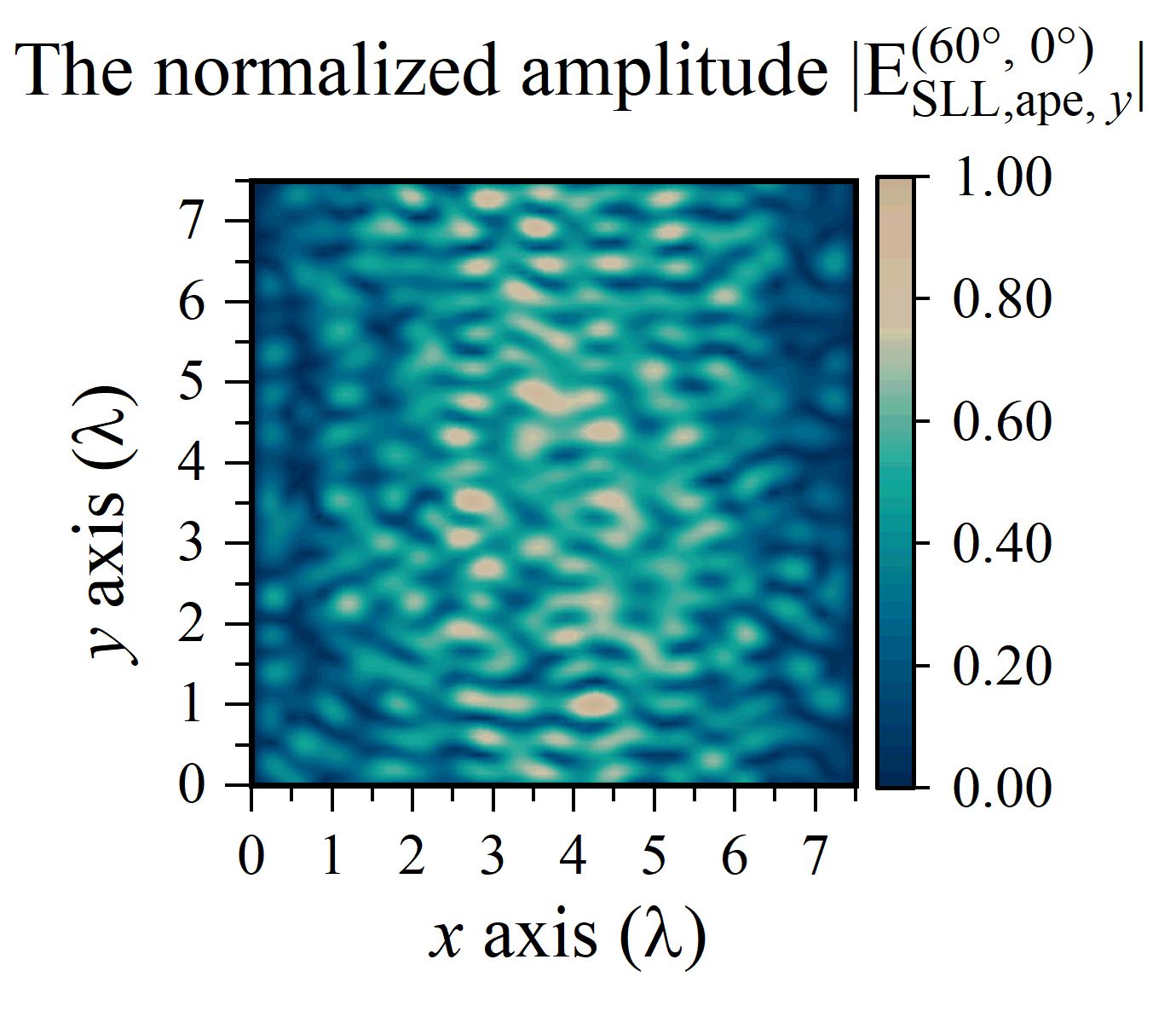}}
	\vspace{-1.0 em}
	\\
	\subfloat[]{\includegraphics[width=0.42\columnwidth]{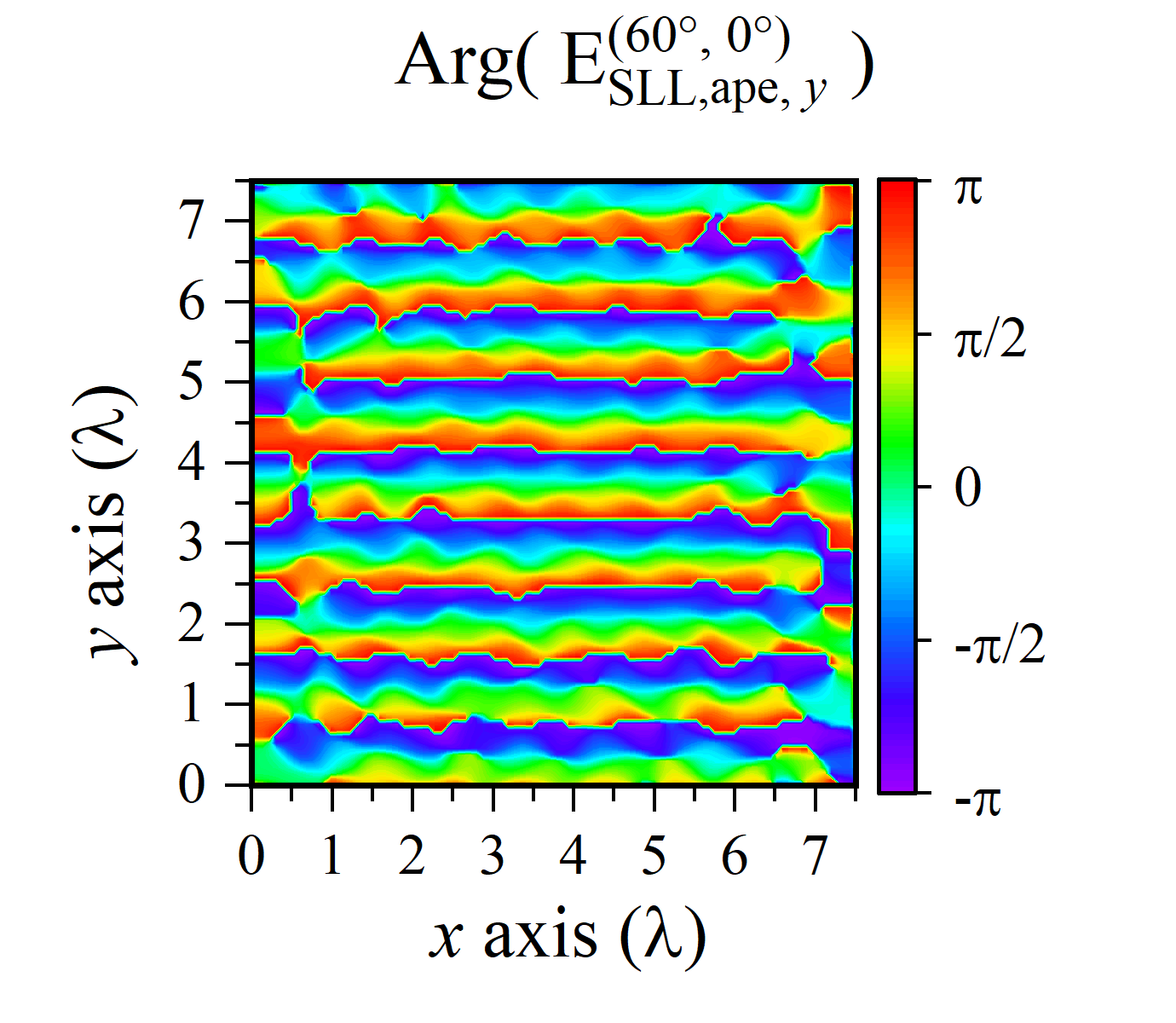}}
	\subfloat[]{\includegraphics[width=0.42\columnwidth]{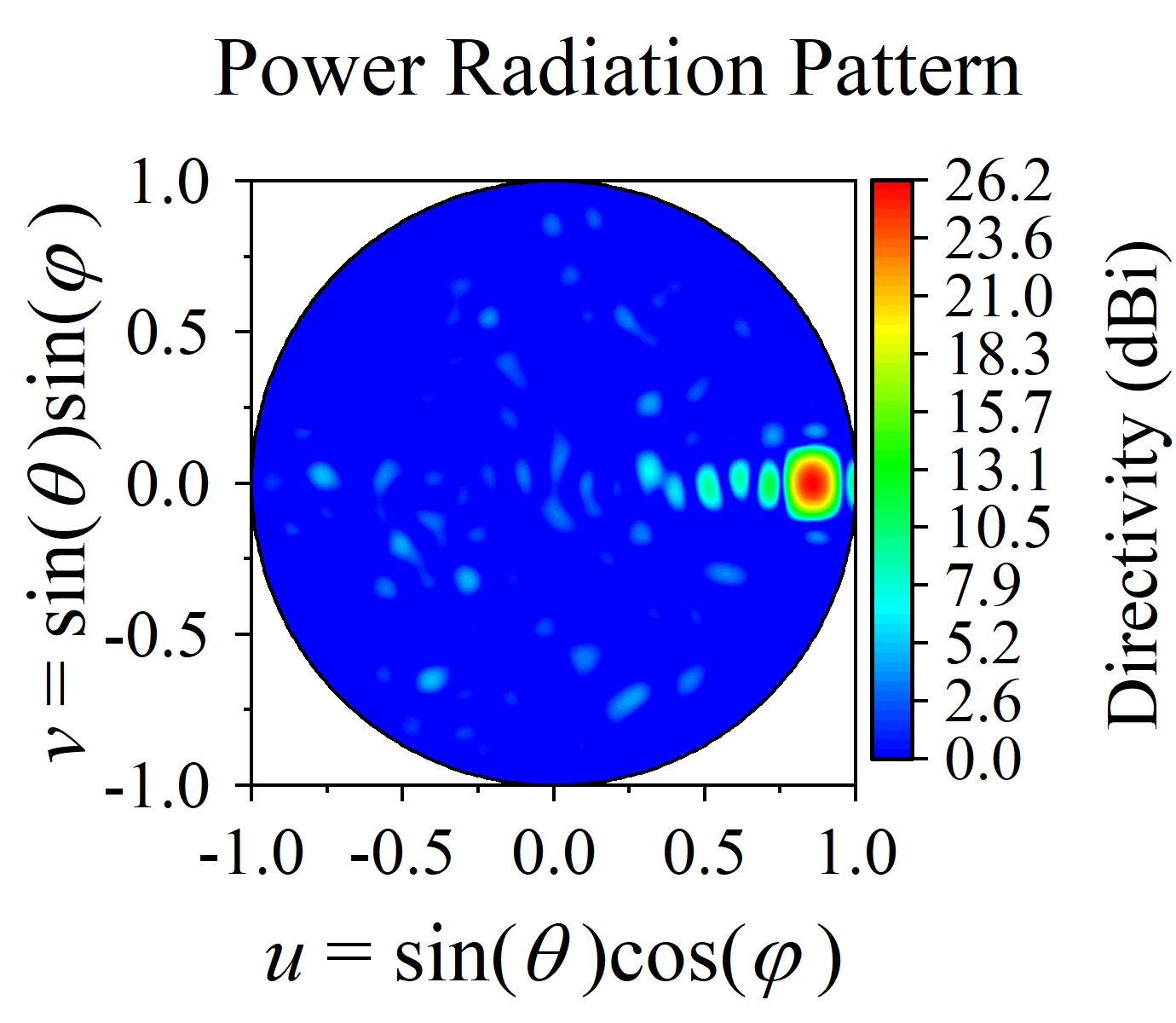}}
	\caption{
		(a) Target pattern with low SLL;
		(b) $\left| E_{\mathrm{SLL, ape},y}^{(60^\circ,0^\circ)} \right|$;
		(c) Arg $\left( E_{\mathrm{SLL, ape},y}^{(60^\circ,0^\circ)} \right) $;  
		(d) Power radiation pattern of $ E_{\mathrm{SLL, ape},y}^{(60^\circ,0^\circ)} $. 
	}
	\label{fig_sll}
\end{figure}

The side lobe level(SLL) in Fig. \ref{fig_CSTcalfarfield}(d) and in Fig. \ref{fig_dualportArray} are not good. 
This is caused by using Fig. \ref{fig_dipole} as the target, rather than what IDMBSA brings. 
%图\ref{fig_CSTcalfarfield}(d)，以及图\ref{fig_dualportArray}中的sll，并不令人满意。这是我们以\ref{fig_dipole}为目标导致的，而非IDMBSA带来的。
To handle this situation, simply use the ideal pattern like Fig. \ref{fig_sll}(a) as the target. 
The polarization constraint remains \eqref{eq_polar}. 
%极化约束仍然为式\eqref{eq_polar}
%The results are shown in Fig. \ref{fig_sll}(b, c, d). 
The aperture field obtained by IDMBSA is denoted as $E_{\mathrm{SLL, ape},y}^{(60^\circ,0^\circ)}$ and shown in Fig. \ref{fig_sll}(b, c). 
%想要改善这一情况也非常简单，若我们以图x为目标方向图, 利用IDMBSA得到的口径场分布记作 $E_{\mathrm{SLL, ape},y}^{(60^\circ,0^\circ)}$
%\begin{CJK}{UTF8}{gbsn} 
%\end{CJK}

\subsubsection{Large Aperture}
Furthermore, IDMBSA remains applicable for large apertures. 
We once attempted to design a pencil beam pointing at 60 degrees on a super-sized aperture of 50 $ \lambda $ and achieved excellent results. During this design process, we observed that a lower number of modes $M$ and $N$ made it difficult for the cost function to converge. However, once the number of modes exceeds a certain critical value (roughly 40), the cost function suddenly converges. 
This is an interesting phenomenon, and we believe it may hold significant implications for researching in performance limits. 

%\begin{CJK}{UTF8}{gbsn} 
%此外，IDMBSA方法在特大口径仍然适用。我们曾尝试在具有50个波长的超大口径上设计一个指向60°的pencil beam，并取得了出色的结果。在这一设计过程中，我们发现，对于口径为50波长的情况，较低的模式数$M$和$N$会导致目标函数难以下降；然而，随着模式数超过某一临界值，优化过程会突然变得异常顺利。这是一个非常有趣的现象，我们认为这可能对研究性能上限具有重要意义！
%\end{CJK}

\subsubsection{Transmitarray Antenna}
%IDMBSA can not only be applied to array synthesis and sparse array design but also to the design of transmitting arrays or reflecting arrays. 
IDMBSA can also be applied to the design of transmitting arrays.
In such cases, it is necessary to consider the relative phase between the feeding antenna and the array elements and make appropriate adjustments. Detailed discussions and experimental validations can be found in the work of the authors \cite{icmmt_self}, and we do not delve into further details here.

%离散化后的IDMBSA的口径场除了可以应用于阵列综合和稀布阵中，也应用于透射阵或反射阵的设计，此时需要考虑馈源天线与阵面单元的相对相位并做出修正，具体的实验验证，在作者[引用ICMMT论文]有详细的讨论，这里不再赘述
In addition, as mentioned in the Introduction, the aperture field distribution obtained by IDMBSA can also be used as a design objective for existing aperture field implementation methods \cite{TensorMetasurfaceGetArbitraryAperture} and \cite{CavityGetArbitraryAperture}.
By combining aperture field implementation methods and IDMBSA, it is possible to achieve the design of radiating devices with arbitrary far-field radiation pattern.
%此外，正如我们在引言中提到的IDMBSA得到的口径场分布也可以直接作为\cite{TensorMetasurfaceGetArbitraryAperture}和\cite{CavityGetArbitraryAperture} 的设计目标。将上述两篇文章的设计思路结合本文提出的IDMBSA，可直接实现对任意远场辐射方向图的辐射器件设计。
\section{Conclusion}
In this communication, we propose a method called IDMBSA for designing aperture field from arbitrary phaseless radiation pattern. The resulting aperture field obtained through IDMBSA, after discretization, can be used for traditional array synthesis. By applying suitable thresholding, it can  be applied to sparse array design. 
Compared to traditional array synthesis methods and sparse array design methods, IDMBSA significantly reduces the dimensionality of the solution space. Furthermore, the utilization of analytical solutions further reduces the computational burden. 
Moreover, IDMBSA enables the independent design of dual-polarization radiation patterns. 
We have validated these through numerical and full-wave simulations, which show excellent agreement with the expected outcomes. 
Additionally, the smoothness of the aperture field obtained by IDMBSA makes it suitable as a design objective for existing aperture field implementation methods such as \cite{TensorMetasurfaceGetArbitraryAperture} and \cite{CavityGetArbitraryAperture}. By combining these methods, it is possible to achieve the direct design of radiating devices from arbitrary phaseless far-field radiation pattern.
However, there is still much follow-up work to be done, such as determining the limits of achievable radiation pattern complexity with finite apertures. These research will contribute to further advancing the application of IDMBSA in electromagnetic device design.

%综上所述，本文提出了一种名为IDMBSA的方法，用于从任意无相位模式设计孔径场。通过IDMBSA得到的口径场在离散化后可用作传统阵列综合的设计输出，并可以直接应用于稀疏阵列设计，通过适当的阈值筛选。
%IDMBSA相比与传统的阵列综合方法和稀疏阵设计方法，我们的优化变量大幅度减少，降低了解空间的维度，利用解析解还可以进一步降低计算负担。
%此外，在特定情况下，IDMBSA还能够实现双极化辐射方向图的独立设计。通过数值仿真和全波仿真验证，我们发现仿真结果与预期完全吻合。
%上述是IDMBSA优势的一部分，此外IDMBSA的口径场天然具有平滑性，可以作为现有某些器件设计方法的设计目标，结合这些方法，我们能够根据任意辐射方向图的要求设计出满足需求的电磁器件
%然而，仍然有许多后续工作需要进行，例如确定有限孔径下可实现的辐射方向图复杂性的界限等，这将有助于进一步推动IDMBSA在电磁器件设计中的应用。

\appendices
\section{Analytical expression of Equation \eqref{eq10}}\label{sec:appd-a}
Equation \eqref{eq10_ana} is the analytic expression of \eqref{eq10}.
\begin{subequations}\label{eq10_ana}
	\begin{equation}
		\begin{aligned}
			|f_x(\theta ,\phi )|=
			|\sum_{n=1}^{N}
			&
			\left( 
			\sum_{m=1}^{M}\alpha_{mn}^x \prod_{i=1}^3{g_{1,i}}\left( m,n,\theta ,\phi \right)
			\right.
			\\
			& \left. + \sum_{m=0}^{M}\beta_{mn}^{x}
			\prod_{i=1}^3{g_{2,i}}\left( m,n,\theta ,\phi \right)
			\right)|
		\end{aligned}
	\end{equation}
	\begin{equation}\label{eq10b_ana}
		\begin{aligned}
			|f_y(\theta ,\phi )|=
			|\sum_{m=1}^{M}
			&
			\left(
			\sum_{n=1}^{N}\alpha_{mn}^y \prod_{i=1}^3{g_{1,i}}\left( m,n,\theta ,\phi \right)
			\right.
			\\
			&\left. + \sum_{n=0}^{N}\beta_{mn}^{y}
			\prod_{i=1}^3{g_{3,i}}\left( m,n,\theta ,\phi \right)
			\right)|
		\end{aligned}
	\end{equation}
\end{subequations}
where  $ g_{j,i} $ are defined in \eqref{g_ji}.
\begin{small}
\begin{subequations}\label{g_ji}
\begin{align}
		&\chi _1=\sin \theta \sin \phi 
		\\
		&\chi _2=\sin \theta \cos \phi 
\\
\begin{split}
		&g_{1,1}=
		\left( 4abe^{-\frac{1}{2}jk\cos \phi \left( b\cos \theta +a\sin \theta \right)} \right) 
		\\
		&\quad \times \Big[ \left( 4n^2\pi ^2+b^2k^2\left( \cos \left( 2\phi \right) +2\cos \left( 2\theta \right) \sin ^2\left( \phi \right) -1 \right) \right)\Big.
		\\
		&\quad \Big. \times \left(  a^2 k^2 \chi_2^2-m^2\pi ^2 \right) \Big] ^{-1}
\end{split}
\\
&g_{1,2}=\left( -m\pi +e^{jka\chi _2}\left( m\pi \cos \left( m\pi \right) -jak\sin \left( m\pi \right) \chi _2 \right) \right) 
%\end{align}
%\end{subequations}
%\end{small}
%
%
%\begin{small}
%\begin{subequations}
%\begin{align}
\\
\begin{split} 
&g_{1,3}=\left( e^{\frac{1}{2}jkb\cos \left( \phi +\theta \right)}n\pi \right. 
\\
&\qquad\quad \left. + e^{\frac{1}{2}jkb\cos \left( \phi -\theta \right)}\left( -n\pi \cos \left( n\pi \right) +jbk\sin \left( n\pi \right) \chi _1 \right) \right) 
\end{split}
\\
&g_{3,1}=g_{2,1}=g_{1,1} \tag{16f}
%\end{align}
% \end{subequations}
%\begin{small} % eq_16
%\begin{align}
\\
&g_{2,2}=\left( -ak\chi _2+e^{jka \chi _2}\left( -jm\pi \sin \left( m\pi \right) +ak\cos \left( m\pi \right) \chi _2 \right) \right)\tag{16g}
\\
&g_{2,3}=-jg_{1,3} \tag{16h}
\\
&g_{3,2}=\left( -jm\pi +e^{jka\chi _2}\left( jm\pi \cos \left( m\pi \right) +ak\sin \left( m\pi \right) \chi _2 \right) \right) \tag{16i}
\\
\begin{split}
&g_{3,3}=\left( -bk\chi _1e^{\frac{1}{2}jkb\cos \left( \phi +\theta \right)} \right. 
\\
&\qquad\quad\ \left. + e^{\frac{1}{2}jkb\cos \left( \phi -\theta \right)}\left( -jn\pi \sin \left( n\pi \right) +bk\cos \left( n\pi \right) \chi _1 \right) \right) 
\end{split}\tag{16j}
\end{align}
%\end{small}
\end{subequations}
\end{small}

%\bibliography{Citations}


\end{document}